\documentclass[11pt]{article}

\usepackage{color,graphicx}
\usepackage{young}
\usepackage[vcentermath]{youngtab}
\usepackage{amsmath,amssymb,graphicx,multirow}
\usepackage{hyperref}
\definecolor{darkred}{rgb}{0.65,0.15,0}
\hypersetup{pdfborder={0 0 0},colorlinks=true,urlcolor=darkred,citecolor=blue,linkcolor=darkred,linktocpage=true}

\usepackage{cite}
\usepackage{amsmath}
\usepackage{amsfonts}
\usepackage{amssymb}
\usepackage{graphicx}%
\usepackage{amsthm}
\usepackage{mathrsfs}
\usepackage[T1]{fontenc}
\usepackage{enumerate}
\usepackage{graphicx}%
\usepackage{multirow}%
\usepackage{amsmath,amssymb,amsfonts}%
\usepackage{amsthm}%
\usepackage{mathrsfs}%
\usepackage[title]{appendix}%
\usepackage{xcolor}%
\usepackage{textcomp}%
\usepackage{manyfoot}%
\usepackage{booktabs}%
\usepackage{algorithm}%
\usepackage{algorithmicx}%
\usepackage{algpseudocode}%
\usepackage{listings}%

\definecolor{darkred}{rgb}{0.65,0.15,0}
\hypersetup{pdfborder={0 0 0},colorlinks=true,urlcolor=darkred,citecolor=blue,linkcolor=darkred,linktocpage=true}

\usepackage{cite}
\usepackage{amsmath}
\usepackage{amsfonts}
\usepackage{amssymb}
\usepackage{graphicx}%
\usepackage{amsthm}
\usepackage{mathrsfs}
\usepackage[T1]{fontenc}
\usepackage{enumerate}
\usepackage{graphicx}
\usepackage{multirow}
\usepackage{amsmath,amssymb,amsfonts}
\usepackage{amsthm}
\usepackage{mathrsfs}
\usepackage[title]{appendix}
\usepackage{xcolor}
\usepackage{textcomp}
\usepackage{manyfoot}
\usepackage{booktabs}
\usepackage{algorithm}
\usepackage{algorithmicx}
\usepackage{algpseudocode}
\usepackage{listings}

\usepackage{color}
\theoremstyle{thmstyleone}%
\theoremstyle{thmstyletwo}%

\theoremstyle{thmstylethree}%
\def\4diml{four-dimensional}

\def\-1{^{-1}}

\newcommand{\A}{\mathscr{A}}

\newcommand{\M}{\mathscr{M}}
\newcommand{\D}{\mathscr{D}}
\newcommand{\G}{\mathscr{G}}

\makeatletter

\@addtoreset{equation}{section}
\makeatother
\begin{document}

\thispagestyle{empty}

\vspace{5mm}

\begin{center}
{\LARGE \bf Poisson-Lie T-duality and D-brane Configurations \\[2mm] of $AdS_2 \times H^2 \times H^2$ Backgrounds}

\vspace{11mm}

\normalsize
{\bf Ali Eghbali}\footnote{ORCID iD:~ https://orcid.org/0000-0001-6076-2179}

\vspace{2mm}
{\small  Department of Physics, Faculty of Basic Sciences,\\
Azarbaijan Shahid Madani University,\\ 53714-161, Tabriz, Iran}\\
\vspace{6mm}
\verb"eghbali978@gmail.com; a.eghbali@azaruniv.ac.ir"\\
\vspace{1mm}

\abstract{We construct a non-Abelian dual pair for the $AdS_2 \times H^2 \times H^2$ background via Poisson-Lie T-duality on a semi-Abelian double.
Using a specific parameterization of the six-dimensional Lie group ${A}_2 \times {A}_2 \times A_2$,
we formulate the original $\sigma$-model incorporating the $AdS_2 \times H^2 \times H^2$ metric with a vanishing Kalb-Ramond field ($B$-field), and
derive its dual background, which features a non-vanishing $B$-field and a physical singularity.
We investigate the conformal properties and test the generalized supergravity equations for the complete geometry;
although a generalized Killing vector exists, the full background fails to satisfy these equations.
We then analyze the asymptotic structure and coordinate domains,
revealing distinct exterior and interior sectors separated by curvature-singular boundaries,
alongside an asymptotic approach to the original background.
Finally, we investigate the transformation of $D$-brane boundary conditions under Poisson-Lie T-duality for
the semi-Abelian Drinfeld double  $({\A}_2 \oplus {\A}_2  \oplus {\A}_2 , 6{\A}_1)$.
By analyzing seven classes of gluing matrices, we examine the eigenvalue spectra, identify candidate dual D-brane configurations, and construct the corresponding Dirichlet and Neumann projectors.
We also investigate the integrability of the Neumann distributions and determine the associated gauge-invariant two-forms and world-volume field strengths, providing a systematic framework for studying the geometric and boundary condition properties of the dual branes.}
\medskip

\end{center}
\setcounter{page}{1}

\newpage
\tableofcontents

\section{Introduction}
\label{Sec.I}
Recently, a number of supersymmetric $AdS_6$ black hole solutions has been constructed by
six-dimensional $F(4)$ gauged supergravity coupled to four vector multiplets with
$ISO(3) \times U(1)$ gauge group, in such a way that the solutions interpolate between the supersymmetric
$AdS_6$ vacuum and near horizon geometries of the form $AdS_2 \times {\cal M}_4$, where ${\cal M}_4$ is considered to be
$ H^2 \times H^2$ and $ H^2 \times S^2$ with $ S^2$ and $H^2$
being a two-sphere and a two-dimensional hyperbolic space, respectively \cite{parinya1}  (see, also, \cite{parinya2}).
In Ref. \cite{Minwoo.Suh1}, a study of supersymmetric $AdS_6$ black holes has been carried out in matter coupled $F(4)$ gauged supergravity.
There, after deriving the supersymmetry equations, the $AdS_2$ solutions, which were first found in \cite{S.M.Hosseini}, are obtained.
Then, it is shown that the $AdS_2$ horizon exists only for the $ H^2 \times H^2$ background and not for the $ H^2 \times S^2$
or $ S^2 \times S^2$ backgrounds (see, also,  \cite{Minwoo.Suh2}).

In the present work, we investigate the  $AdS_2 \times H^2 \times H^2$ geometry and construct a non-Abelian T-dual pair for this background.
Target space duality (T-duality) is one of the most important duality symmetries in string theory.
It connects seemingly different backgrounds in which
the strings can propagate, in such a way that string backgrounds and their dual fields can be considered
as different descriptions of the same physical system.
Klimcik and Severa proposed Poisson-Lie T-duality \cite{Klim1,Klim2} as a generalization of Abelian T-duality \cite{Buscher} and
traditional non-Abelian duality \cite{delaossa}.
They showed that Poisson-Lie T-dual $\sigma$-models are defined by Poisson-Lie group manifolds which constitute a Drinfeld double \cite{Drinfeld}, and then
proved the phase space equivalence of a model and its Poisson-Lie dual pair.

First of all, we investigate the geometric structure and the conformal invariance conditions of the $\sigma$-model with the $AdS_2 \times H^2 \times H^2$ metric up to two-loop order.
After establishing the relevant isometry subgroup and their role in the background symmetry,
we show that, in the absence of a $B$-field and for a constant dilaton, the vanishing of the beta-function equations at the considered order formally requires the parameter relations $\alpha' = l^2=k$ and $\Lambda={3}/{2l^2}$.
However, these relations imply that the dimensionless expansion parameters satisfy $\alpha'/ l^2= \alpha'/k=1$,
rendering the perturbative expansion uncontrolled.
Therefore, the resulting solution should be interpreted as a formal solution of the truncated two-loop beta-function equations rather than as a controlled perturbative demonstration of conformal invariance.
Then, after a review of the construction of Poisson-Lie T-dual $\sigma$-models on the Lie groups, we arrive at the main purposes of our work.

In this paper, our goal is twofold:
first, by applying the Poisson-Lie T-duality approach to a semi-Abelian double, we construct a non-Abelian dual pair for the $AdS_2 \times H^2 \times H^2$ geometry. In particular, using a specific parametrization of the six-dimensional non-Abelian Lie group ${A}_2 \times {A}_2 \times {A}_2$,
we derive the original $\sigma$-model, which recovers the $AdS_2 \times H^2 \times H^2$ metric in the absence of a $B$-field.
We then demonstrate that the dual background, constructed via Poisson-Lie T-duality, is supported by a non-vanishing $B$-field
and a metric whose components become ill-defined in the regions ${\tilde x}_2 = \pm l^2$.
We also investigate whether the non-unimodularity of the dualized algebra  ${\A}_2 \oplus {\A}_2 \oplus {\A}_2$
leads to a generalized supergravity background. Although the metric-invariance condition admits a generalized Killing vector, a direct analysis shows that the remaining generalized supergravity equations are not satisfied for the complete dual geometry.
This highlights that non-unimodularity alone is insufficient to guarantee generalized supergravity,
particularly when the underlying $\sigma$-model is not formulated in the Green-Schwarz framework.
An important aspect of the dual background is the analysis of its global coordinate structure and asymptotic behavior.
Since the dual metric contains a non-trivial denominator, its coordinate domain separates into distinct regions according to its sign,
and the parametrization adopted for the exterior region does not cover the entire dual target space.
We therefore analyze the exterior and interior regions separately and examine their causal and curvature properties.
In particular, the asymptotic behavior reveals an important connection between the dual and original geometries:
at large values of the radial coordinates, the scalar curvature of the dual metric approaches the constant curvature of
the $AdS_2 \times H^2 \times H^2$ metric. Moreover, after an appropriate asymptotic coordinate transformation,
the dual metric approaches precisely the $AdS_2 \times H^2 \times H^2$ background. Thus, although the full dual background possesses a distinct global structure, it asymptotically recovers the original geometry and can be regarded as an asymptotically $AdS_2 \times H^2 \times H^2$ background.

Another important result of this work concerns the transformation of $D$-brane boundary conditions under Poisson-Lie T-duality,
discussed in Section \ref{Sec.V}.
Our strategy here is to follow a prescription invented by Albertsson {\it et al.} \cite{CC}.
Using the duality map obtained from the canonical transformation description of the Poisson-Lie T-duality \cite{sfetsos,sfetsos1}, we
derive the duality transformation of the gluing matrix for the semi-Abelian Drinfeld double $({\A}_2 \oplus {\A}_2  \oplus {\A}_2 , 6{\A}_1)$
and apply it to seven classes of $D$-brane configurations.
The analysis combines the exact eigenvalue spectra of the dual gluing matrices with the construction of Dirichlet and Neumann projectors and the examination of the integrability of the Neumann distributions. We further determine the gauge-invariant two-forms and the corresponding world-volume field strengths, thereby providing a systematic characterization of the candidate dual brane configurations and their boundary condition properties.

It is worth noting that one of the main approaches to study of $D$-branes in various type II string backgrounds \cite{ooguri}
is the boundary state formalism \cite{Boundary.states}.
This approach showed that $D$-branes probe a new aspect of the background geometry, namely the geometry of submanifolds.
Using the boundary state approach, the relation of exact $D$-brane configurations in the Nappi-Witten
background with T-duality transformations was described in \cite{Stanciu1}.
Then, the problem of determining possible $D$-branes in the Nappi-Witten background was revisited
and it was shown that in addition to the known branes, there are also $D$-instantons, flat Euclidean $D$-strings and curved $D$-membranes that admit
parallel spinors \cite{Stanciu2}.
For further study on the subject of $D$-branes in WZW models, refer to \cite{D.brane.WZW} (see, also, \cite{eghbali11}).
On can also find a full set of boundary
conditions compatible with $N = 1$ superconformal symmetry in \cite{lind1} which was analyzed
by studying requirements for invariance of the action, and by
studying the conserved supercurrent (see, also, \cite{lind2}).
In addition, a derivation of the most general local boundary conditions necessary for T-duality
to be compatible with superconformal invariance of the two-dimensional $N = 1$ supersymmetric $\sigma$-model  with boundaries, was given in \cite{lind3}.
There, the behavior of the boundary conditions under T-duality,
and the interpretation of the results in terms of D-branes have been investigated.
Following this work, in \cite{Snobl}, the conditions for open strings with charged endpoints in the
language of gluing matrices, and the D-branes invariance under the
Poisson-Lie T-plurality were studied. The effects of non-Abelian T-duality on $D$-brane supergravity solutions,
including the resulting brane configurations and their charges, have also been investigated in the literature \cite{Tsimpis}.

The structure of the paper is as follows.
Following the Introduction, Section \ref{Sec.II} investigates the geometric structure and the conformal invariance conditions of the
$AdS_2 \times H^2 \times H^2$ $\sigma$-model up to two-loop order.
Section \ref{Sec.III} reviews the construction of Poisson-Lie T-dual $\sigma$-models on the Lie groups, where necessary formulas are summarized.
Section \ref{Sec.IV} contains one of the original results of the work:
using the formulation introduced in Section \ref{Sec.III} and
then starting from the semi-Abelian double $({\A}_2 \oplus {\A}_2  \oplus {\A}_2 , 6{\A}_1)$, we construct
a pair of Poisson-Lie T-dual $\sigma$-models including the $AdS_2 \times H^2 \times H^2$ metric and its dual pair.
Investigation of the structure of the dual geometry
including the singularity and asymptotic nature are discussed at the end of Section \ref{Sec.IV}.
Section \ref{Sec.V} is devoted to the study of $D$-brane configurations under Poisson-Lie T-duality.
We first review the worldsheet boundary conditions and present the algebraic transformation law relating the gluing matrices of the original and dual $\sigma$-models. We then investigate the resulting dual $D$-brane configurations by calculating the exact eigenvalue spectra of the dual gluing matrices, constructing the corresponding Dirichlet and Neumann projectors, and examining the integrability conditions of the Neumann distributions.
We also discuss the gauge-invariant world-volume two-form, defined in terms of the pullback of the dual $B$-field and the world-volume field strength,
and analyze its dependence on the different coordinate branches.
Finally, in Section \ref{Sec.VI}, we present our conclusions and sketch possible future developments.
Detailed derivations regarding the gauge invariance of the world-volume two-form are provided in Appendix \ref{app:A},
while the variational derivation of the Neumann condition and the gluing block is deferred to Appendix \ref{app:B}.
Furthermore, the investigation of $D$-brane structures and the relevant formulae for boundary conditions in the Lie algebra frame are left to Appendix \ref{app.C}.


\section{ Isometries and conformal invariance of the $AdS_2 \times H^2 \times H^2$ background}
\label{Sec.II}

The line element associated to the  $AdS_2 \times H^2 \times H^2$ space is given by
\begin{eqnarray}
ds^2 = \frac{l^2}{r^2} (-d t^2 + d r^2) + k \Big[\frac{1}{y^2} (d x^2 + d y^2) + \frac{1}{u^2} (d v^2 + d u^2)\Big],\label{2.1}
\end{eqnarray}
where $(t , r)$ stand for the coordinates of $AdS_2$, while $(x , y)$ and $(u, v)$ denote two copies of the coordinates of the hyperbolic space $H^2$.
Here, $l$ denotes the $AdS_2$ radius and $k>0$ is the scale parameter associated with the two hyperbolic factors. We consider
the coordinate domain
\begin{equation}
\M=\left\{(t,r,x,y,u,v)\in\mathbb{R}^6:r>0,\quad y>0,\quad u>0\right\}.\label{2.1.1}
\end{equation}
The scalar curvature of this product geometry is constant and is given by ${\cal R} =-2/l^2 -4/k$.
Thus, although the scalar curvature is constant, the geometry is not flat, since each of its three factors has non-vanishing curvature.

In order to investigate the non-Abelian T-duality of the metric \eqref{2.1} within the Poisson-Lie T-duality framework, we seek an isometry subgroup
acting freely and transitively on the target manifold \cite{Nakahara}. Such a group action allows the target space to be identified with the corresponding Lie
group, thereby providing the geometric setting required for the construction of a Poisson-Lie T-dual $\sigma$-model without spectator coordinates.
The $AdS_2$ part of the metric admits three independent Killing vectors,
\begin{align}\label{2.2}
{\cal \xi}_{_+}=\frac{1}{2} (t^2+ r^2) \partial_{t} +t r \partial_{r}, ~~~~~{\cal \xi}=t\partial_{t}+ r\partial_{r},~~~~~{\cal \xi}_{_-}=- \partial_{t}.
\end{align}
The isometry Lie algebra spanned by these vectors is the $\mathfrak{sl}(2,\mathbb R)$ with the commutation relations:
$
[{\cal \xi} , {\cal \xi}_{_+}]={\cal \xi}_{_+}, [{\cal \xi} , {\cal \xi}_{_-}]=- {\cal \xi}_{_-}, [{\cal \xi}_{_+} , {\cal \xi}_{_-}]={\cal \xi}.
$
This algebra admits two possible two-dimensional subalgebras,
namely those spanned by $({\cal \xi}, {\cal \xi}_{_+})$ and $({\cal \xi}, {\cal \xi}_{_-})$.
Both are algebraically isomorphic to the two-dimensional non-Abelian Lie algebra, denoted by ${\A}_2$;
however, this raises the question of whether either choice is equally suitable for constructing the desired group action.
Although the subalgebra spanned by $({\cal \xi}, {\cal \xi}_{_+})$ is algebraically valid,
its generators become linearly dependent along the curves $t =\pm r$, where ${\cal \xi}_{_+}$ is also null.
Consequently, this choice does not generate a free and transitive action throughout the $AdS_2$ Poincar\'{e} patch.
To investigate this, we consider the determinants of the matrices formed by the components of their generators with respect to the coordinates $(t, r)$.
For the first choice, $({\cal \xi}, {\cal \xi}_{_+})$, the determinant is ${r}(t^2-r^2)/2$.
This determinant vanishes along the curves $t =\pm r$, where the two Killing vectors become linearly dependent.
Moreover, the squared norm of ${\cal \xi}_{_+}$ is $-l^2(t^2-r^2)^2/(4r^2)$, showing that it becomes null precisely on these curves.
Thus, despite being a valid two-dimensional Lie subalgebra algebraically, the pair $({\cal \xi}, {\cal \xi}_{_+})$
fails to remain linearly independent throughout the Poincar\'{e} patch and consequently does not generate a free and transitive action there.
For the second choice, $({\cal \xi}, {\cal \xi}_{_-})$, the corresponding determinant is $r$.
Since $r > 0$ throughout the Poincar\'{e} patch, this determinant never vanishes,
and the two generators remain linearly independent everywhere in that domain.
In this case, the generators induce a free and transitive action on the Poincar\'{e} half-plane.
Furthermore, ${\cal \xi}_{_-}$ is timelike everywhere, with squared norm $-l^2/r^2$.
Therefore, while both choices are equivalent at the level of Lie algebra structure,
the latter provides a globally regular realization of the two-dimensional subgroup on the Poincar\'{e} patch.

For $H^2$ in the  $(x , y)$ coordinate system, one finds the following three independent Killing vectors
\begin{align}\label{2.3}
{\cal N}_{_+}=(x^2-y^2) \partial_{x} + 2 x y \partial_{y}, ~~~~~{\cal N}= x \partial_{x}+ y \partial_{y},~~~~~{\cal N}_{_-}=- \partial_{x}.
\end{align}
These Killing vectors also satisfy in the $\mathfrak{sl}(2,\mathbb R)$ algebra.
The squared norms of the three Killing vectors associated with the $H^2$ sector are strictly positive throughout the upper half-plane, provided that $k>0$.
Hence, all three generators are everywhere spacelike and non-vanishing.
However, the causal character of individual Killing vectors alone does not guarantee the regularity of the corresponding group action.
To examine the linear independence of a pair of generators, one must evaluate the determinant of their component matrix.
For the pair $({\cal N}, {\cal N}_{_+})$, this determinant is $y(x^2+y^2)>0$, where $y>0$ throughout $H^2$.
In particular, along $x=0$, the two generators reduce to $\mathcal N=y\partial_y$ and $\mathcal N_+=-y^2\partial_x$, respectively,
and remain linearly independent. Thus, the pair $({\cal N}, {\cal N}_{_+})$ spans the tangent space at every point of $H^2$ and
induces a transitive action of the corresponding connected Lie group.
Similarly, for the alternative pair $({\cal N}, {\cal N}_{_-})$, the determinant of the component matrix is $y>0$.
Consequently, this pair is also linearly independent everywhere on $H^2$.
For the second $H^2$ factor, whose coordinates are $(u,v)$, the corresponding Killing vectors
are obtained by interchanging $x\leftrightarrow v$ and $y\leftrightarrow u$:
\begin{align}\label{2.3.5}
\widehat{\mathcal N}_{+} =(v^2-u^2)\partial_v+2uv\partial_u,~~~~~
\widehat{\mathcal N} =v\partial_v+u\partial_u, \qquad \widehat{\mathcal N}_{-}=-\partial_v.
\end{align}
In particular, the pair $(\widehat{\mathcal N},\widehat{\mathcal N}_{-})$
is linearly independent throughout $u>0$, since the determinant of its component matrix in the coordinates $(u,v)$ is $u>0$.
Since the metric \eqref{2.1} is a direct product of three two-dimensional constant-curvature geometries, its continuous isometry Lie algebra is
\begin{equation}\label{2.3.6}
\mathfrak{iso}(\M)=\mathfrak{sl}(2,\mathbb R) \oplus\mathfrak{sl}(2,\mathbb R)\oplus\mathfrak{sl}(2,\mathbb R).
\end{equation}
Indeed, each of the three two-dimensional factors admits three independent Killing vectors, and the Killing vectors belonging to different factors commute.

We now construct an explicit six-dimensional isometry subalgebra of $\mathfrak{iso}(\M)$ that is isomorphic to $\A_2\oplus \A_2\oplus \A_2$.
For this purpose, we choose the following generators:
\begin{align}
T_1^{(0)}=-\mathcal{\xi},~~~T_2^{(0)}&=\mathcal{\xi}_{-},~~~
T_1^{(1)}=-\mathcal N,~~~T_2^{(1)}=\mathcal N_{-},~~~
T_1^{(2)}=-\widehat{\mathcal N},~~~T_2^{(2)}=\widehat{\mathcal N}_{-}.\label{2.3.7}
\end{align}
These generators satisfy the following non-vanishing commutation relations
\begin{equation}\label{2.3.8}
[T_1^{(i)},T_2^{(i)}]=T_2^{(i)}, \qquad i=0,1,2.
\end{equation}
Consequently, the six generators span the decomposable Lie algebra $\G \simeq \A_2\oplus \A_2\oplus \A_2$.
We next examine the finite action of the corresponding Lie group.
We define a full group element by
\begin{equation}\label{2.3.9}
g=g_{_0}(s_0,\tau_0)  g_{_1}(s_1,\tau_1) g_{_2}(s_2,\tau_2).
\end{equation}
where
\begin{equation}\label{2.3.10}
g_{_i}(s_i,\tau_i) = e^{\tau_i T_2^{(i)}}e^{s_iT_1^{(i)}}, \qquad i=0,1,2.
\end{equation}
Here, $s_i$ and $\tau_i$ are independent real group parameters; they parametrize the finite group transformations and should not be
confused with the coordinates of the target manifold.
Integrating the flows generated by the six Killing vectors in \eqref{2.3.7}, the action of $g$ on the coordinates $(t,r,x,y,u,v)$ is obtained as
\begin{equation}\label{2.3.11}
(t,r,x,y,u,v) \longrightarrow \left(e^{-s_0}t-\tau_0,  e^{-s_0}r, e^{-s_1}x-\tau_1, e^{-s_1}y, e^{-s_2}u, e^{-s_2}v-\tau_2\right)
\end{equation}
We now prove that this group action is free. Suppose that a group element leaves a point $(t,r,x,y,u,v)\in \M$ invariant. From Eq.~ \eqref{2.3.11}, we obtain
\begin{equation}
\begin{aligned}\label{2.3.12}
e^{-s_0}r&=r,        & e^{-s_1}y&=y,          & e^{-s_2}u&=u, \\
e^{-s_0}t-\tau_0&=t, & e^{-s_1}x-\tau_1&=x,   & e^{-s_2}v-\tau_2&=v.
\end{aligned}
\end{equation}
Since $r>0$, $y>0$, and $u>0$, the first three equations imply $s_0=s_1=s_2=0$. Substituting these values into the remaining
equations gives $\tau_0=\tau_1=\tau_2=0$. Thus, the only group element that leaves any point of $\M$ invariant is the identity element. Therefore, the action is free.

To establish transitivity, consider two arbitrary points of the target manifold, $ p=(t_1,r_1,x_1,\\$$y_1,u_1,v_1)$ and
$q=(t_2,r_2,x_2,y_2,u_2,v_2)$. We seek a group element $g$ such that $g\cdot p=q$. This equation is satisfied by choosing
\begin{align}\label{2.3.13}
s_0&=\ln\big(\frac{r_1}{r_2}\big), & \tau_0&=\frac{r_2}{r_1}t_1-t_2, \nonumber\\
s_1&=\ln\big(\frac{y_1}{y_2}\big), & \tau_1&=\frac{y_2}{y_1}x_1-x_2, \nonumber\\
s_2&=\ln\big(\frac{u_1}{u_2}\big), & \tau_2&=\frac{u_2}{u_1}v_1-v_2.
\end{align}
Because $r_1,r_2,y_1,y_2,u_1,u_2$ are strictly positive, all six parameters are well-defined real numbers.
Hence, for any two points in $\M$, there exists a group element
that maps one to the other. This proves that the action of the subgroup $A_2 \times A_2 \times A_2$
is transitive on the target space defined by the metric \eqref{2.1}.
We have thus explicitly shown that the six-dimensional isometry subgroup $A_2 \times A_2 \times A_2$ acts freely and transitively on the target
manifold defined by the metric \eqref{2.1}. Since the dimension of this group equals the dimension of the target manifold, no spectator
coordinates are required for this group action. This provides the geometric basis for the non-Abelian T-duality construction of the
background \eqref{2.1}, which will be carried out in Section~\ref{Sec.IV}.

Before closing this section, let us have a look at the vanishing of the two-loop beta-function equations with the $AdS_2 \times H^2 \times H^2$ metric.
According to \cite{callan}, the conformal invariance condition of the $\sigma$-model is provided by the vanishing of the beta-function equations.
One can find a presentation of the beta-function equations at two-loop level in \cite{c.hull,Metsaev}.
Solving the vanishing of the two-loop beta-function equations for the G\"{o}del universe,
Barrow and D\c{a}browski \cite{Barrow} found a simple relation between the angular velocity of
the G\"{o}del universe, $\Omega$, and the inverse string tension of the form $\alpha' =1/\Omega^2$ in the absence of $B$-field.
Subsequently, a notable result concerning the G\"{o}del metric and its T-duals was presented in \cite{Godel}.
In the present work, by employing the two-loop beta-function equations established in \cite{Godel} (see also \cite{Eghbali.2,ERA2,egh.fortsch}),
we investigate the conformal invariance conditions for the metric \eqref{2.1} up to two-loop order.
In the absence of a $B$-field (where the field strength $H =d B$ vanishes), the conditions for the vanishing of the two-loop beta-function are given by
\begin{align}
&{\cal R}_{{\mu \nu}}+{\nabla}_{\mu} {\nabla}_{\nu} \Phi +\frac{1}{2} \alpha' {\cal R}_{{\mu  \rho\sigma\eta}} {\cal R}_{\nu}^{{~\rho\sigma\eta}}
+{\cal O}(\alpha'^2) = 0, \label{2.4}\\
&2 \Lambda + {\nabla}^2 \Phi - ({\nabla} \Phi)^2
-\alpha' \frac{1}{4} {\cal R}_{{\mu\nu\rho\sigma}} {\cal R}^{{\mu\nu\rho\sigma}} +{\cal O}(\alpha'^2)=0,\label{2.5}
\end{align}
where ${\cal R}_{{\mu \nu}}$ and $ {\cal R}_{{\mu \nu \rho \sigma}} $ are the Ricci tensor and
Riemann tensor field of the metric $G_{{\mu \nu}}$, respectively. Also, $\Phi$ denotes the dilaton field.
For the metric \eqref{2.1}, the non-vanishing components of the Ricci tensor and the Kretschmann scalar are found to be
${\cal R}_{{tt}} =-{\cal R}_{{rr}} =1/r^2,  {\cal R}_{{xx}} ={\cal R}_{{yy}} =-1/y^2, {\cal R}_{{vv}} ={\cal R}_{{uu}} =-1/u^2$, and
${\cal K} = {\cal R}_{{\mu\nu\rho\sigma}} {\cal R}^{{\mu\nu\rho\sigma}} =4/l^4 + 8/k^2$, respectively.
Substituting these expressions into the beta-function equations \eqref{2.4} and \eqref{2.5},
with a vanishing $B$-field and a constant dilaton, yields the formal conditions $\alpha' = l^2=k$ and $\Lambda={3}/{2l^2}$
for the vanishing of the equations at the displayed order.
However, the resulting parameter relations imply $\alpha'/ l^2= \alpha'/k=1$, indicating that the curvature expansion is not perturbatively controlled.
In particular, the higher-order contributions in the $\alpha'$-expansion are not guaranteed to be negligible, and the cancellation between the leading and two-loop terms cannot be regarded as a controlled perturbative result. This issue is analogous to the one discussed in Ref. \cite{ERA2} for the
$AdS_2$ background, where the higher-order corrections likewise render the $\alpha'$-expansion uncontrollable. Therefore, the present calculation establishes only a formal solution of the beta-function equations at the truncated order; it does not, by itself, demonstrate perturbative or exact conformal invariance of the $AdS_2 \times H^2 \times H^2$ $\sigma$-model.


\section{ A brief review of Poisson-Lie T-duality}
\label{Sec.III}

In this section, we briefly review the construction of Poisson-Lie T-dual $\sigma$-models on Lie groups
\cite{Klim1,Klim2}. Let $\Sigma$ be a two-dimensional worldsheet mapped into a $d$-dimensional target space
manifold $\M$, equipped with a metric $G_{\mu\nu}$ and an antisymmetric tensor field
$B_{\mu\nu}$. In light-cone coordinates $\sigma^\pm=(\tau\pm\sigma)/2$, the corresponding $\sigma$-model action is written as
\begin{eqnarray}
S= \frac{1}{2}\int_{_{\Sigma}}\!d\sigma^+  d\sigma^-
\big({G}_{{\mu\nu}}(X)+{B}_{{\mu\nu}}(X)\big)  \partial_{_+} X^\mu \partial_{_-} X^{\nu},\label{3.1}
\end{eqnarray}
where $\partial_\pm:=\partial/\partial\sigma^\pm = \partial_\tau \pm \partial_\sigma$, and $X^\mu$, $\mu=1,\ldots,d$, are the target space coordinates.

Suppose that a Lie group $G$ acts freely on the target manifold $\M$ from the right. The infinitesimal
generators of this action are described by vector fields $V_a=V_a^\mu\partial_\mu$, satisfying $[V_a,V_b]=f^c{}_{ab}V_c$,
where $f^c{}_{ab}$ are the structure constants of the Lie algebra $\G$ of $G$. Unlike an ordinary
isometry symmetry, Poisson-Lie symmetry does not require the background tensor to be invariant under the group
action. Instead, the Lie derivative of ${\mathcal E}_{\mu\nu}=G_{\mu\nu}+B_{\mu\nu}$ is governed
by the structure constants of a dual Lie algebra $\widetilde{\G}$, with $\dim\widetilde \G=\dim G$. The corresponding
Poisson-Lie dualizability condition is \cite{Klim1,Klim2}
\begin{eqnarray}\label{3.2}
{\cal L}_{_{V_{_a}}}{\cal E}_{{\mu\nu}} = {{\widetilde f}^{bc}}_{~~a}~{V_{b}}^{{~\sigma}}~ {V_{c}}^{{\rho}}~{\cal E}_{{\mu \rho}}~
{\cal E}_{{\sigma \nu}},
\end{eqnarray}
where ${{\widetilde f}^{bc}}_{~~a}$ are the structure constants of the dual Lie algebra $\tilde \G$ of $\tilde G$.
The compatibility of the Poisson-Lie condition \eqref{3.2} with the Lie brackets of the generating
vector fields imposes the following mixed Jacobi identities:
\begin{eqnarray}\label{3.3}
{f^a}_{bc}{\widetilde{f}^{de}}_{\; \; \; \; a}=
{f^d}_{ac}{\widetilde{f}^{ae}}_{\; \; \; \;  b} +
{f^e}_{ba}{\widetilde{f}^{da}}_{\; \; \; \;  c}+
{f^d}_{ba}{\widetilde{f}^{ae}}_{\; \; \; \; c}+
{f^e}_{ac}{\widetilde{f}^{da}}_{\; \; \; \; b}.
\end{eqnarray}
These compatibility conditions define a Lie bialgebra structure and lead naturally to the Drinfeld double $\D$.
A Drinfeld double is a Lie algebra equipped with a non-degenerate, symmetric, ad-invariant bilinear form $\langle\cdot,\cdot\rangle$
and a decomposition ${\D}=\G \oplus \widetilde{\G}$ as vector spaces, where $\G$ and $\widetilde{\G}$ are maximally isotropic
Lie subalgebras. The triple $(\D,\G,\widetilde{\G})$ is called a Manin triple.
We choose a basis in each of the subalgebras as
$T_{{_a}} \in \G$  and ${\widetilde T}^{a} \in {\widetilde \G}, a = 1, ...,$ dim G, such that
\begin{eqnarray}\label{3.4}
\big<T_{{_a}} ,  T_{{_b}}\big> = \big<{\widetilde T}^{{^a}} ,  {\widetilde T}^{{^b}}\big> =0,~~~~~~~\big<T_{{_a}} , {\widetilde T}^{{^b}}\big>  = {\delta}_{_{a}}^{{~b}}.
\end{eqnarray}
Furthermore, the basis of the two subalgebras satisfy the commutation relations
\begin{eqnarray}\label{3.5}
[T_a , T_b] = {f^{c}}_{ab} ~T_c,~~~~~[{\widetilde T}^{a} , {\widetilde T}^{b}] ={{\widetilde f}^{ab}}_{~~c} ~{\widetilde T}^{c},~~~~
[T_a , {\widetilde T}^{b}] = {{\widetilde f}^{bc}}_{~~a} {T}_c + {f^{b}}_{ca} ~{\widetilde T}^{c}.
\end{eqnarray}
The Drinfeld double provides the common algebraic structure underlying the original and dual Poisson-Lie $\sigma$-models \cite{Klim1,Klim2}.

We now specialize to the case in which the group $G$ acts freely and transitively on the target manifold
$\M$. After choosing a reference point, the target space can then be identified with the group
manifold, $\M \simeq G$. In this case, the $\sigma$-model can be formulated directly in terms of the group-valued field $g$, with the action
\begin{eqnarray}\label{3.6}
S=\frac{1}{2}\int\!d\sigma^{+} d\sigma^{-}  {{R_{_+}}}^{\hspace{-1.5mm} a}\;{{R_{_-}}}^{\hspace{-1.5mm} b} {E_{_{ab}}}(g),
\end{eqnarray}
where $R_{\pm}^a$ are the components of the right-invariant Maurer-Cartan one-forms which are constructed by means of
an element $g$ of ${G}$ in the following way
\begin{eqnarray}\label{3.7}
R_{\pm}^a := (\partial_{_\pm} g~g^{-1})^a = \partial_{_\pm} X^{\mu}~ R_{\mu}^{~a},
\end{eqnarray}
and ${E_{_{ab}}}(g)$ defined by
\begin{eqnarray}\label{3.8}
 {E_{_{ab}}}(g) = \big(E_{_0}^{-1}(e) + \Pi (g)\big)^{-1},
\end{eqnarray}
is a certain bilinear form on $G$. Here, $E_0(e)$ is a constant and invertible matrix evaluated at the identity element $e$ of the Lie group $G$. It
specifies the constant coupling matrix of the $\sigma$-model at the identity, while the dependence of $E_{ab}(g)$ on
the group element $g$ is encoded in the Poisson-Lie structure $\Pi(g)$. The latter is constructed from the
adjoint action of $G$ on the Drinfeld double algebra $\D$. In particular, the adjoint action can be decomposed as
\begin{eqnarray}\label{3.10}
g^{-1} T_{{a}}~ g &=&a_{_{a}}^{{~c}}(g) ~ T_{{c}},\nonumber\\
g^{-1} {\widetilde T}^{{a}} g &=& b^{{ac}}(g)~ T_{{c}}+(a^{-1})_c^{~a}(g)~{\widetilde T}^{{c}},
\end{eqnarray}
where $a(g)$ and $b(g)$ are the corresponding blocks of the adjoint representation matrix. The Poisson-Lie structure is then defined by
\begin{eqnarray}\label{3.9}
\Pi^{ab}(g)=b^{ac}(g) ~ (a^{-1})_c^{~b}(g),
\end{eqnarray}
which is an antisymmetric bivector and vanishes at the identity element, $\Pi(e)=0$. Consequently, $E_{ab}(g=e)=E_{0\,ab}(e)$.

The Poisson-Lie dual $\sigma$-model is obtained by exchanging the two maximally isotropic subgroups,
$G\leftrightarrow\widetilde G$, and their Lie algebras. The corresponding coupling matrix and Poisson structure are transformed according to
$E_0\leftrightarrow\widetilde E_0=E_0^{-1}$ and $\Pi(g)\leftrightarrow\widetilde\Pi(\widetilde g)$.
Since the two groups have equal dimensions, the dual model is described by a group-valued field $\widetilde g\in\widetilde G$ and has the action
\begin{eqnarray}\label{3.11}
\tilde S = \frac{1}{2} \int d\sigma^{+}  d\sigma^{-}~ {\widetilde R}_{+_{a}}{\widetilde R}_{-_{b}} {{{\widetilde E}}^{{ab}}}(\widetilde g),
\end{eqnarray}
where
\begin{eqnarray}\label{3.12}
{{\widetilde E}}(\widetilde g) = \big(E_{0}(e)+ {\widetilde \Pi}(\widetilde g)\big)^{-1}.
\end{eqnarray}
Here, ${\widetilde R}_{\pm_{a}}$ denote the components of the right-invariant one-forms on ${\widetilde G}$,
defined by $(\partial_{_\pm} {\widetilde g}~{\widetilde g}^{-1})_a={\widetilde R}_{{\pm}_a}=\partial_{_\pm} {\widetilde X}^{\mu} {\widetilde R}_{\mu a}$, where
${\widetilde X}^{\mu}$ $({\mu}= 1, ..., \dim {\widetilde G})$ parameterize an element $\widetilde g$ of $\widetilde G$.
The dual Poisson structure $\widetilde\Pi(\widetilde g)$ is defined analogously through the corresponding
adjoint-action matrices of the Drinfeld double.
The actions \eqref{3.6} and \eqref{3.11} therefore describe a pair of Poisson-Lie T-dual $\sigma$-models \cite{Klim1,Klim2}.

Finally, we consider the semi-Abelian Drinfeld double, which corresponds to non-Abelian T-duality. In this case, the original Lie algebra $\G$ is
non-Abelian, whereas its dual algebra $\widetilde{\G}$ is Abelian. Thus, ${f^a}_{bc} \neq 0$ and $\widetilde f^{ab}_{~~c}=0$.
Consequently, the Poisson structures on the original and dual groups reduce to
\begin{equation}\label{3.13}
\Pi^{ab}(g)=0,~~~~~~~~~~~{\tilde \Pi}_{ab}(\tilde g) = -{\tilde x}_{{c}}\; {f^c}_{ab},
\end{equation}
where ${\tilde x}_{{a}}$'s are local coordinates characterizing the group element $\tilde g$. We have in this case,
$ E=E_{0}(e)$ and $\widetilde E(\widetilde g) = \left(E_0(e)-\widetilde x_c {\cal Y}^c\right)^{-1}$ where $({\cal Y}^c)_{ab} = f^c{}_{ab}$, and thus the action
\eqref{3.6} takes the following form
\begin{eqnarray}\label{3.14}
S=\frac{1}{2}\int\!d\sigma^{+} d\sigma^{-}  {{R_{_+}}}^{\hspace{-1.5mm} a}\;{{R_{_-}}}^{\hspace{-1.5mm} b} {E_{_0}}_{ab}(e).
\end{eqnarray}
Thus, the original $\sigma$-model reduces to the standard group $\sigma$-model with the constant coupling matrix $E_0(e)$.
If $E_0(e)$ is symmetric, the antisymmetric part of the background tensor vanishes and hence the original $B$-field is zero. Conversely, an antisymmetric part
of $E_0(e)$ gives rise to a non-vanishing $B$-field.
In the next section, we employ the semi-Abelian Drinfeld double to construct the non-Abelian T-dual of
the $AdS_2 \times H^2 \times H^2$ background introduced in Section~\ref{Sec.II}.


\section{The non-Abelian T-dual of the $AdS_2 \times H^2 \times H^2 $ background}
\label{Sec.IV}

In this section, we explicitly construct a pair of Poisson-Lie T-dual $\sigma$-models associated with a
semi-Abelian Drinfeld double. The original and dual target spaces are six-dimensional group manifolds,
$\M \simeq G$ and $\widetilde{\M} \simeq \widetilde G$, respectively.
Here, $G=A_2\times A_2\times A_2$ is the six-dimensional Lie group whose Lie algebra is $\A_2 \oplus \A_2\oplus \A_2$.
As established in Section \ref{Sec.II}, this group acts freely and transitively on the original target
manifold. The dual group $\widetilde G$ is taken to be the six-dimensional Abelian Lie group $6A_1$.
We first show that the original $\sigma$-model reproduces the $AdS_2\times H^2\times H^2$ geometry.
We then construct the corresponding non-Abelian T-dual background and investigate its geometric and curvature properties.

\subsection{ Original background as the $AdS_2 \times H^2 \times H^2$ metric}
\label{Sec.IV.1}

Let us start with the construction of the Drinfeld double following from the Lie
algebra ${\A}_2 \oplus {\A}_2  \oplus {\A}_2$ and its dual pair, $6{\A}_1$.
The Manin triple $\big(\D , {\A}_2 \oplus {\A}_2  \oplus {\A}_2 , 6{\A}_1\big)$ possesses twelve generators
$(T_{_a}, {\tilde T}^a)$, a=1, ..., 6,  so that they obey the following set of commutation relations
\begin{align}\label{4.1}
[T_{_1} , T_{_2}] =& T_2,~~~~~~~~[T_{_3} , T_{_4}]=T_{_4},~~~~~~~~~[T_{_5} , T_{_6}]=T_{_6},\nonumber\\
[T_{_1} , {\widetilde T}^2] =& -{\widetilde T}^2,~~~~[T_{_3} , {\widetilde T}^4]=-{\widetilde T}^4,~~~~
~~[T_{_5} , {\widetilde T}^6]=-{\widetilde T}^6,\nonumber\\
[T_{_2} , {\widetilde T}^2] =& {\widetilde T}^1,~~~~~~~~[T_{_4} , {\widetilde T}^4]={\widetilde T}^3,~~~~~~~~
[T_{_6} , {\widetilde T}^6]={\widetilde T}^5,\nonumber\\
{~~~~~[\widetilde T^a , \widetilde T^b]}=& 0, ~~~~\qquad a, b=1, \ldots, 6.
\end{align}
Note that all the remaining independent commutators vanish.
Here, $(T_{_1}, \cdots, T_{_6})$ and $({\widetilde T}^1, \cdots, {\widetilde T}^6)$ are generators of the Lie algebras $ {\A}_2 \oplus {\A}_2  \oplus {\A}_2$ and
$6{\A}_1$, respectively.
In order to write the action \eqref{3.6} (or \eqref{3.14}) on the group manifold ${A}_2 \times {A}_2 \times {A}_2$
explicitly we need to find the corresponding right-invariant one-forms.
To this purpose we use the following parameterization of the group manifold:
\begin{eqnarray}\label{4.2}
g~=~e^{x_{_1} T_{_1}}~e^{x_{_2} T_{_2}}~e^{x_{_3} T_{_3}}~e^{x_{_4} T_{_4}}~e^{x_{_5} T_{_5}}~e^{x_{_6} T_{_6}},
\end{eqnarray}
where $(x_{_1}, \cdots, x_{_6})$ stand for the group coordinates. Using \eqref{3.7}, \eqref{4.1} and \eqref{4.2} one then gets
\begin{align}\label{4.3}
R_{\pm}^1=& \partial_{\pm} x_{_1},~~~~~~~~R_{\pm}^2 = e^{x_{_1}}~\partial_{\pm} x_{_2},\nonumber\\
{R_{\pm}^3}=& \partial_{\pm} {x}_{_3},~~~~~~~~R_{\pm}^4 = e^{x_{_3}}~\partial_{\pm} { x}_{_4},\nonumber\\
{R_{\pm}^5}=& \partial_{\pm} {x}_{_5},~~~~~~~~R_{\pm}^6 = e^{x_{_5}}~ \partial_{\pm} { x}_{_6}.
\end{align}
As discussed at the end of Section \ref{Sec.III}, when the dual Lie group is considered to be Abelian,
the Poisson structure on $G$ vanishes. So, we have in this example, $\Pi(g) =0$.
To achieve a $\sigma$-model with the $AdS_2 \times H^2 \times H^2$ metric
one should choose the $\sigma$-model constant matrix as follows:
\begin{eqnarray}\label{4.4}
{E_0}(e)=\left( \begin{array}{cccccc}
                    l^2 & 0 & 0 & 0 & 0 & 0\\
                     0 & -l^2 & 0 & 0 & 0 & 0\\
                     0 & 0 & k & 0 & 0 & 0\\
                     0 & 0 & 0 & k & 0 & 0\\
                      0 & 0 & 0 & 0 & k & 0\\
                       0 & 0 & 0 & 0 & 0 & k
                      \end{array} \right),
\end{eqnarray}
where the parameters $l^2$ and $k$ have already been introduced in \eqref{2.1}.
Inserting \eqref{4.3} and \eqref{4.4} into \eqref{3.14},
the original $\sigma$-model is worked out to be
\begin{align}\label{4.6}
S = \frac{1}{2} \int d \sigma^+ d \sigma^- \Big[l^2\big(\partial_+ x_{_1} \partial_- x_{_1} -e^{2 x_{_1}}  \partial_+ x_{_2} \partial_- x_{_2}\big)
&+k\big(\partial_+ x_{_3} \partial_- x_{_3} +e^{2 x_{_3}}  \partial_+ x_{_4} \partial_- x_{_4}\big) \nonumber\\
&+k\big(\partial_+ x_{_5} \partial_- x_{_5} +e^{2 x_{_5}}  \partial_+ x_{_6} \partial_- x_{_6}\big)\Big].
\end{align}
By identifying action \eqref{4.6} with the $\sigma$-model of the form \eqref{3.1}
one concludes that $B$-field is zero, and the corresponding line element is read off to be
\begin{eqnarray}\label{4.7}
ds^2 = l^2 \big(d x_{_1}^2 - e^{2 x_{_1}}~ dx_{_2}^2\big) +k \big(d x_{_3}^2 +e^{2 x_{_3}}~ dx_{_4}^2\big)
+k \big(d x_{_5}^2 + e^{2 x_{_5}}~ dx_{_6}^2\big).
\end{eqnarray}
Now, one may use the coordinate transformation
\begin{align}\label{4.7.1}
r &= e^{-x_1}, \qquad t=x_2, \nonumber\\
y &= e^{-x_3}, \qquad x=x_4, \nonumber\\
u &= e^{-x_5}, \qquad v=x_6,
\end{align}
then, the resulting metric turns into the $AdS_2 \times H^2 \times H^2$ one, as shown in Eq. \eqref{2.1}.
Thus, we have constructed a non-Abelian T-dual $\sigma$-model on the six-dimensional group manifold ${A}_2 \times {A}_2 \times {A}_2$
whose background describes the $AdS_2 \times H^2 \times H^2$ metric only. Below,
we construct the non-Abelian T-dual space of this metric.

\subsection{ Dual background}
\label{Sec.IV.2}

In order to construct the dual $\sigma$-model on the group manifold $6A_1$
we parameterize the group with the coordinates ${\tilde x}_{_{a}} = ({\tilde x}_{_{1}}, \cdots, {\tilde x}_{_{6}})$ so that its element
is defined as \eqref{4.2} by replacing untilded quantities with tilded ones.
Then, we find that ${\tilde R}_{{\pm}_a}=\partial_{_\pm} {\tilde x}_{_{a}}$.
Also, utilizing the commutation relations of \eqref{4.1} and formula \eqref{3.13} one gets
\begin{eqnarray}\label{4.8}
{\tilde \Pi}(\tilde g) = {\tilde \Pi}^{^{(12)}} \oplus {\tilde \Pi}^{^{(34)}}  \oplus {\tilde \Pi}^{^{(56)}},
\end{eqnarray}
where submatrices ${\tilde \Pi}^{{(ij)}}$ are defined as follows:
\begin{eqnarray}\label{4.9}
{\tilde \Pi}^{{(ij)}} = \left( \begin{array}{cc}
                    0 & -{\tilde x}_{_j} \\
                     {\tilde x}_{_j} & 0 \\
                      \end{array} \right), \qquad i=1, 3, 5,~ j=i+1.
\end{eqnarray}
The dual background matrix can be obtained by inserting \eqref{4.8} and ${E_0}(e)$ of \eqref{4.4} into \eqref{3.12}.
It is then read off
\begin{align} \label{4.10}
{{{\tilde E}}}(\tilde g) ={ \widetilde E}^{(12)} \oplus {\widetilde E}^{(34)} \oplus {\widetilde E}^{(56)},
\end{align}
where
\begin{align}\label{4.10.1}
\widetilde E^{(12)}=\frac{1}{\Delta_{12}}
\begin{pmatrix}
l^2&-\widetilde x_2\\
\widetilde x_2 &-l^2
\end{pmatrix}, \qquad
\widetilde E^{(ij)}  =\frac{1}{\Delta_{ij}}
\begin{pmatrix}
k & \widetilde x_j\\
-\widetilde x_j & k
\end{pmatrix}, \qquad i=3, 5,~ j=i+1,
\end{align}
where $\Delta_{_{12}} =l^4 - {\tilde x}_{_2}^2, \Delta_{_{34}} =k^2 +{\tilde x}_{_4}^2$, and $\Delta_{_{56}} =k^2 + {\tilde x}_{_6}^2$.
Putting these pieces together into \eqref{3.11}, one can obtain
the action of dual $\sigma$-model. Then, comparing
the resulting action with the $\sigma$-model of the form \eqref{3.1},
the dual metric and ${\tilde B}$-field take the following forms
\begin{align}
{\tilde ds}^2 =& \frac{l^2}{\Delta_{_{12}}} \big(d {\tilde x_{_1}}^2 - d {\tilde x_{_2}}^2\big)
 + \frac{k}{\Delta_{_{34}}} \big(d {\tilde x_{_3}}^2 + d {\tilde x_{_4}}^2\big) + \frac{k}{\Delta_{_{56}}} \big(d {\tilde x_{_5}}^2+ d {\tilde x_{_6}}^2\big),\label{4.11}\\
{\tilde  B }=& -\frac{{\tilde x}_{_2}}{\Delta_{_{12}}} d {\tilde x_{_1}}\wedge d {\tilde x_{_2}}+
\frac{{\tilde x}_{_4}}{\Delta_{_{34}}} d {\tilde x_{_3}}\wedge d {\tilde x_{_4}}+\frac{{\tilde x}_{_6}}{\Delta_{_{56}}} d {\tilde x_{_5}}\wedge d {\tilde x_{_6}}.\label{4.12}
\end{align}
It is worth noting that, although the dual background has a non-vanishing $\widetilde B$-field, its field
strength vanishes identically, $\widetilde H = d\widetilde B=0$.
In fact, on the regular domain defined by $\Delta_{12}\neq 0$, the two-form $\widetilde B$ is exact and can be
written as $\widetilde B=d\widetilde\Lambda$, where
\begin{align}\label{4.12.1}
\widetilde\Lambda=-\frac{1}{2}\bigg[\ln|\Delta_{12}|\,d\widetilde x_1 +\ln(\Delta_{34})\,d\widetilde x_3
+\ln(\Delta_{56})\,d\widetilde x_5\bigg].
\end{align}
Thus, the non-vanishing $\widetilde B$-field does not generate a non-trivial three-form field strength in this background.
For $k>0$, the functions $\Delta_{34}$ and $\Delta_{56}$ are strictly positive for real coordinates.
Hence, the only potentially singular locus of the dual metric is determined by
$\Delta_{12}=0$, or equivalently, $\widetilde x_2=\pm l^2$. To determine whether these
singularities are true curvature singularities or merely coordinate singularities, we evaluate the scalar curvature of the dual geometry.
However, prior to this detailed analysis, it is instructive to first establish the
conformal invariance and generalized supergravity equations of the dual background.

\subsubsection{Conformal invariance and generalized supergravity equations of the dual background}

It is natural to investigate whether the dual background defined by equations \eqref{4.11} and \eqref{4.12}
can satisfy the one-loop Weyl-invariance conditions of the ordinary $\sigma$-model, or,
alternatively, the corresponding generalized supergravity equations.
We first recall that the Poisson-Lie dilaton transformation \cite{vonUnge} provides a relation between the dilaton fields of the original and dual
$\sigma$-models. However, interpreting this transformation as a map between one-loop conformal $\sigma$-models requires the original background,
including its dilaton, to satisfy the corresponding one-loop Weyl-invariance conditions.
In the present case, the original $AdS_2\times H^2\times H^2$
background considered in Section \ref{Sec.II}, with vanishing $B$-field and constant dilaton, does not satisfy these conditions.
More explicitly, for $H=0$, the metric beta-function equation at one-loop order takes the form ${\cal R}_{{\mu \nu}}+{\nabla}_{\mu} {\nabla}_{\nu} \Phi=0$.
For a constant dilaton, this condition reduces to ${\cal R}_{{\mu \nu}}=0$, which is not satisfied by the original background.
Indeed, as we already showed in Section \ref{Sec.II}, the Ricci tensor of the original background is non-vanishing,
so we are led to a contradiction: the metric beta-function equation is not fulfilled.
Furthermore, allowing the dilaton to be a general coordinate-dependent function does not restore
one-loop Weyl invariance: our direct analysis shows that the complete set of one-loop beta-function
equations admits no solution for the dilaton. Thus, the failure of one-loop conformal invariance
is not restricted to the constant-dilaton choice considered in Section \ref{Sec.II}.
Consequently, the standard Poisson-Lie dilaton transformation cannot be interpreted here as a
transformation between two one-loop conformal $\sigma$-models.
The dilaton obtained through this formal transformation therefore does not, by itself, establish conformal invariance of the complete dual background.

The situation is further clarified by the algebra involved in the dualization.
The non-vanishing structure constants of the dualized algebra ${\A}_2 \oplus {\A}_2  \oplus {\A}_2$,
obtained from the commutation relations in \eqref{4.1}, are $f^2{}_{12}=f^4{}_{34}=f^6{}_{56}=1$.
Consequently, the trace vector $n_a:=f^b{}_{ab}$ is given by $n_a=(1,0,1,0,1,0)$.
Since this vector is non-vanishing, the Lie algebra ${\A}_2 \oplus {\A}_2  \oplus {\A}_2$
is non-unimodular\footnote{A Lie algebra ${\G}$ is called unimodular if $\operatorname{Tr}(\operatorname{ad}_{T_a})=0$
for every generator $T_a$. With the convention $[T_a,T_b]=f^c{}_{ab}T_c$, this condition is equivalent to $f^b{}_{ab}=0$ for all $a$.}.
Non-unimodularity is known to be closely related to generalized supergravity in non-Abelian T-duality and Yang-Baxter deformations, where the trace of the structure constants is associated with a generalized Killing vector \cite{Arutyunov3,Wulff1,Wulff3.revised}.
For the present semi-Abelian double, the non-vanishing components of the trace vector are associated with
the generators $T_1$, $T_3$, and $T_5$, and hence identify the corresponding directions of the candidate generalized Killing vector in the dual
background. However, this relation should not be interpreted as a purely algebraic criterion
guaranteeing that every non-unimodular dual background necessarily satisfies the generalized
supergravity equations. The complete set of supergravity equations must still be checked explicitly for the resulting background.

For the dual background given in \eqref{4.11} and \eqref{4.12}, we have therefore tested the generalized supergravity equations directly.
In particular, one of the equations requires
\begin{align}
{\cal L}_{_I} {\tilde G}_{\mu \nu} =0,
\label{4.2.1-13}
\end{align}
where $I$ is called the generalized supergravity Killing vector \cite{Arutyunov3}.
Since the dual metric \eqref{4.11} is independent of $\widetilde x_1$, $\widetilde x_3$, and
$\widetilde x_5$, it admits the three Killing vectors
$\partial_{\widetilde x_1}$, $\partial_{\widetilde x_3}$, and $\partial_{\widetilde x_5}$.
We therefore consider their general constant linear combination,
\begin{align}
{I}= c_{_1} \partial_{{\tilde x_1}} + c_{_2} \partial_{{\tilde x_3}}+ c_{_3} \partial_{{\tilde x_5}},
\label{4.2.1-14}
\end{align}
where $ c_{_1},  c_{_2},  c_{_3}$ are arbitrary constants.
To test the complete set of generalized supergravity equations, one must also specify the generalized one-form ${\mathbb {X}}_\mu$.
In the conventions adopted here, it is related to the one-form $Z_\mu$ through ${\mathbb X}_\mu=I_\mu+Z_\mu$,
where $I_\mu=\widetilde G_{\mu\nu}I^\nu$, and $Z_\mu=\partial_\mu\Phi+\widetilde B_{\nu\mu}I^\nu$ \cite{Arutyunov3}.
In our analysis, we allow the dilaton field $\Phi$ to be a completely general coordinate-dependent scalar function,
without imposing any particular functional form or restricting its coordinate dependence. Thus,
we consider the most general $Z_\mu$ generated by the above parametrization for the candidate
Killing vector \eqref{4.2.1-14}.
Substituting this expression, together with the dual metric \eqref{4.11} and the $\widetilde B$-field \eqref{4.12}, into the complete generalized supergravity
equations, we find that the resulting system admits no solution, even for a general coordinate-dependent dilaton $\Phi$.
The failure of the dual background to satisfy the generalized supergravity equations does not contradict the established relation between
non-unimodular duality transformations and generalized supergravity.
This relation is derived within specific string-theoretic frameworks, particularly the Green--Schwarz formulation,
and its applicability relies on the underlying structure of the corresponding string $\sigma$-model.
In the present construction, the original $AdS_2\times H^2\times H^2$ model is a purely
bosonic $\sigma$-model and has not been formulated as a Green-Schwarz superstring theory.
Therefore, the standard Green-Schwarz-based arguments relating non-unimodularity to generalized supergravity do not automatically imply that
the dual background must satisfy the generalized supergravity equations.

There is an important conceptual reason why this result is not in conflict with the standard relation between
non-unimodular deformations and generalized supergravity.
In several well-known examples, the connection between non-unimodular Yang-Baxter deformations and generalized supergravity is established within the Green-Schwarz formulation of string theory.
For example, the analysis of the $\eta$- and $\lambda$-deformed supercoset string models in Ref. \cite{Wulff1} relies on the Green-Schwarz structure and, in particular, on the associated kappa symmetry.
Similarly, the systematic derivation of non-Abelian T-duality transformations for Green-Schwarz strings in
Ref. \cite{Wulff3.revised} establishes the corresponding supergravity interpretation within that framework.
Thus, the appearance of a generalized Killing vector for a non-unimodular deformation is tied to a specific string-theoretic setting and cannot be transferred automatically to an arbitrary bosonic $\sigma$-model.
This distinction is further illustrated by studies of OSP$(1|2)$ $\sigma$-models.
In Ref. \cite{samadi}, the non-unimodular Yang--Baxter-deformed OSP$(1|2)$ WZW backgrounds
were found not to satisfy the corresponding graded generalized supergravity equations,
whereas the undeformed background admits a solution with an appropriate supervector field.
The analysis shows that non-unimodularity alone does not guarantee the existence of a suitable
generalized one-form required by these equations. Similarly, the super non-Abelian T-duality
analysis of the OSP$(1|2)$ principal chiral model in Ref. \cite{Bielli1} showed that the dual
background does not satisfy the corresponding three-dimensional supergravity constraints.
These examples illustrate that the supergravity properties of a dual background depend not only
on the algebraic structure of the duality transformation but also on the underlying $\sigma$-model and its formulation.

\subsubsection{The structure and asymptotic behavior of the dual geometry}
\label{Sec.IV.3}

The dual metric \eqref{4.11} contains the denominator
\begin{align}
\Delta_{12}=l^4-{\tilde{x}_2}^{2}.\label{4.13}
\end{align}
Therefore, the coordinate domain of the dual geometry separates into three sectors: the exterior region,
$|{\tilde{x}_2}|>l^2$, where $\Delta_{12} <0$; the interior region, $|{\tilde{x}_2}|<l^2$, where $\Delta_{12}>0$;
and the boundary surfaces ${\tilde{x}_2}=\pm l^2$, where $\Delta_{12}$=0.
We analyze the exterior and interior regions separately below.
\\
\\
{\bf Exterior region: $|\tilde{x}_2|>l^2$}
\\
To investigate the structure and asymptotic behavior of the dual geometry in the exterior region, we first introduce the coordinate transformation
\begin{align}
{\tilde x_{_1}} = & l^2 t,    &{\tilde x_{_2}} = l^2 \cosh r,~~~~~~~~~~~  \nonumber\\
{\tilde x_{_3}} = & k x,      &{\tilde x_{_4}} = k \sinh y,~~~~~~~~~~~~ \nonumber\\
{\tilde x_{_5}} = & k v,      &{\tilde x_{_6}} = k \sinh u,~~~~~~~~~~~~
\label{4.14}
\end{align}
where $r > 0$. This parametrization covers the positive exterior branch, $\tilde{x}_2 \geq l^2$;
the negative exterior branch, $\tilde{x}_2 < -l^2$, can be described analogously by taking $\tilde{x}_2=-l^2\cosh r$.
For both the positive and negative branches, the dual metric becomes
\begin{align}
{\tilde ds}^2 = {l^2} \big(d r^2 - \frac{d t^2}{ \sinh^2 r}\big)
 + k \big(d y^2 + \frac{d x^2}{ \cosh^2 y}\big) + k \big(d u^2 + \frac{d v^2}{ \cosh^2 u}\big),\label{4.15}
\end{align}
and the corresponding $\tilde{B}$-field is
\begin{align}
{\tilde  B }= {l^2} \coth r~ d t \wedge d r+ k \tanh y~ d x \wedge d y +k \tanh u ~d v \wedge d u.\label{4.16}
\end{align}
As expected from the exactness of the $\tilde{B}$-field discussed above, its field strength vanishes identically, $\widetilde H = d \widetilde B=0$.
Moreover, since $\cosh y\geq 1$ and $\cosh u\geq 1$, the metric components in the $(x,y)$ and $(u,v)$ parts remain regular for finite values of these coordinates. The potentially singular behavior is therefore associated with the $(t,r)$ part. The scalar curvature of the dual metric is
\begin{eqnarray}\label{4.17}
\tilde {\cal R} = -\frac{1}{l^2}-\frac{1}{k}-\frac{3+\cosh^2 r}{l^2 \sinh^2 r} + \frac{5-3 \sinh^2 y}{2 k \cosh^2 y}
+ \frac{5-3 \sinh^2 u}{2 k \cosh^2 u}.
\end{eqnarray}
This expression diverges as $r \to 0$, demonstrating that the surface $r=0$, corresponding to $\tilde{x}_2=l^2$ ($\Delta_{12}$=0),
is a true curvature singularity on this branch. Thus, the singularity cannot be removed by a regular coordinate transformation.
On the other hand, the dual metric is independent of $t, x$, and $v$, and consequently admits the three Killing vectors
$\partial_t$, $\partial_x$, and $\partial_v$.
The vectors $\partial_x$ and $\partial_v$ are spacelike,
whereas the vector $\partial_t$ is timelike throughout the exterior region, with $ \|\partial_t\|^2=\widetilde G_{tt} =-l^2 /{\sinh^2 r} <0$.
Although its norm approaches zero as $r\to\infty$, it does not vanish at any finite positive value of $r$. Hence, there is no finite-radius Killing
horizon in this coordinate patch. Consequently, the curvature singularity at $r=0$ is not hidden
behind a finite-radius Killing horizon in the exterior region.
Thus, the exterior dual geometry retains three commuting Killing isometries, one timelike,
generated by $\partial_t$, and two spacelike, generated by $\partial_x$ and $\partial_v$.

It is also instructive to examine the limiting behavior of the dual background.
In the simultaneous limit $r, y, u \to 0$, using $\sinh r\simeq r$, $\cosh y\simeq 1$, and $\cosh u\simeq 1$,
the metric and $\tilde{B}$-field reduce to
\begin{align}
{\tilde ds}^2 =& {l^2} \big(d r^2 - \frac{d t^2}{r^2}\big)
 + k \big(d y^2 + {d x^2}\big) + k \big(d u^2 + {d v^2}\big),\label{4.18}\\
{\tilde  B }=&  \frac{l^2}{r}~ d t \wedge d r+ k y~ d x \wedge d y +k u ~d v \wedge d u.\label{4.19}
\end{align}
The scalar curvature of the limiting metric behaves as ${\tilde {\cal R}} \simeq -4/(l^2 r^2)$, and therefore diverges at $r=0$.
One might incorrectly infer, from an analysis of the beta-functions associated with the asymptotic dual metric,
that the approximate background given by \eqref{4.18} and \eqref{4.19} along with the dilaton field ${\tilde \Phi}=c_{_0}-2 \log r$ satisfy the one-loop beta-functions equations.
However, because the beta-function depends on second-order derivatives, such an analysis neglects essential contributions, thereby failing to establish the vanishing of the beta-function for the complete dual model.
It is important to emphasize that verifying the one-loop beta-function equations for a truncated
asymptotic background does not establish their vanishing for the complete dual geometry.
The reason is that the beta-function equations contain the Ricci tensor and, at higher order, quadratic curvature terms,
so that sub-leading terms neglected in the metric can contribute at the same or lower order after taking derivatives.
To address this issue, we re-examine the beta-function equations for the complete six-dimensional dual metric \eqref{4.15}.
In the absence of the field strength, ${\tilde H} =d \widetilde{B}=0$, the one-loop metric beta-function has schematically the structure
\begin{align}
\beta_{\mu \nu} [\widetilde{G}_{\mu \nu} , \widetilde{\Phi}] = {\widetilde {\cal R}}_{\mu \nu}[\widetilde{G}]
+ \nabla_\mu \nabla_\nu \widetilde{\Phi} + \cdots.\label{20.1}
\end{align}
Suppose, in general, that the exact metric is decomposed as
\begin{align}
\widetilde{G}_{\mu \nu} = \widetilde{G}_{\mu \nu}^{(0)} + h_{\mu \nu},\label{20.2}
\end{align}
where $\widetilde{G}_{\mu \nu}^{(0)}$ denotes the leading asymptotic metric and $h_{\mu\nu}$
represents the sub-leading terms neglected in the asymptotic approximation. The metric beta-function can then be expressed schematically as
\begin{align}
\beta_{\mu\nu}^{(\widetilde{G})}
[\widetilde{G}^{(0)},\widetilde{\Phi}]
+\delta\beta_{\mu\nu}^{(\widetilde{G})}[h]+\cdots.
\label{20.3}
\end{align}
In particular, the linearized Ricci tensor contains contributions involving second derivatives of the metric correction,
\begin{align}
\delta\widetilde{\cal R}_{\mu\nu} \sim {\nabla} {\nabla} h+\cdots.
\label{20.4}
\end{align}
Therefore, the fact that $h_{\mu\nu} \to 0$ in an asymptotic limit does not, by itself, imply that its second derivatives vanish in the same limit.
We now demonstrate this explicitly for all three coordinate-dependent components of the dual metric \eqref{4.15}.
Expanding the time component of the metric \eqref{4.15} around $r=0$, we obtain
\begin{align}
-\frac{l^2}{ \sinh^2 r} = -\frac{l^2}{r^2} + \frac{l^2}{3} - \frac{l^2 r^2}{15} + {\cal O} (r^4).\label{20.5}
\end{align}
Similarly, expanding the component of the hyperbolic part around $y=0$  yields,
\begin{align}
\frac{k}{\cosh^2y}=k-k y^2+\frac{2k}{3} y^4+{\cal O}(y^6),
\label{20.6}
\end{align}
with an analogous expansion for the $u$-sector.
By comparing the leading asymptotic metric in \eqref{4.18} with the expressions in \eqref{4.15}, \eqref{20.5}, and \eqref{20.6}, we identify the non-zero components of the neglected correction $h_{\mu\nu}$ as follows:
\begin{align}
h_{tt}&=\frac{l^2}{3}
-\frac{l^2r^2}{15} +\mathcal{O}(r^4),\nonumber\\
h_{xx}&=-ky^2+\frac{2k}{3}y^4+\mathcal{O}(y^6), \nonumber\\
h_{vv}&=-ku^2+\frac{2k}{3}u^4+\mathcal{O}(u^6).\label{20.7}
\end{align}
Although these corrections are sub-leading relative to the corresponding leading metric components in
the limit $r, y, u \to 0$, their second derivatives need not vanish. In particular, the relative
correction $h_{tt}/\widetilde G_{tt}^{(0)}$ vanishes as $r\to0$, even though $h_{tt}$ itself approaches the finite value $l^2/3$. Explicitly,
we find
\begin{align}
\partial_r^2 h_{tt}=-\frac{2l^2}{15}+\mathcal{O}(r^2),~~~~
\partial_y^2 h_{xx}=-2k+\mathcal{O}(y^2),~~~~
\partial_u^2 h_{vv}=-2k+\mathcal{O}(u^2).\label{20.8}
\end{align}
Thus, the sub-leading metric corrections possess non-vanishing second derivatives in the asymptotic limit.
Since the linearized Ricci tensor contains terms involving second derivatives of the metric
perturbation, these corrections can contribute to the one-loop metric beta-function even when
they are sub-leading in the metric itself.
This explicitly illustrates why the vanishing of the beta-functions for a leading asymptotic metric, even if established,
cannot by itself demonstrate their vanishing for the complete dual geometry.

In the opposite limit $r, y, u \to \infty$, using $\sinh r \simeq e^r/2$ and $\cosh y \simeq e^y/2$, $\cosh u\simeq e^u/2$,
the dual metric approaches
\begin{align}
{\tilde ds}^2 = {l^2} \big(d r^2 -4 e^{-2 r} {d t^2}\big)
 + k \big(d y^2 + 4 e^{-2 y} {d x^2}\big) + k \big(d u^2 + 4 e^{-2 u} {d v^2}\big).\label{4.20}
\end{align}
The scalar curvature remains finite and approaches the constant value $ -2/l^2 -4/k$ in the simultaneous large-coordinate
limit, which coincides with that of the original $AdS_2 \times H^2 \times H^2$ background.
This behavior is consistent with the properties of asymptotically AdS spaces, where the curvature remains constant at infinity.
Note that if we look more closely at the metric behavior at $r\rightarrow \infty$,
we conclude that the time component of the metric in this region goes to zero, ${\tilde G}_{tt}=-l^2 /{\sinh^2 r}\rightarrow -4l^2 e^{-2 r}\rightarrow 0$,
i.e., the time direction is compressed. This behavior is more like approaching an infinite boundary or a region of extreme redshift,
but the real singularity is at the $r=0$, and this region is a singular boundary of the spacetime, i.e.,
the geometric structure actually breaks down.
Furthermore, introducing the coordinates
\begin{align}
{\tau =2 t},~~~~{\rho =e^{r}}, ~~~~ {\bar{x} =2 x}, ~~~~ {\bar{y} =e^{y}}, ~~~~ {\bar{v} =2 v}, ~~~~{\bar{u} =e^{u}},\label{4.21}
\end{align}
the asymptotic metric takes the form
\begin{align}
{\tilde ds}^2 = \frac{l^2}{\rho^2} \big(d \rho^2 - {d \tau^2}\big)
 + \frac{k}{\bar{y}^2} \big(d \bar{x}^2 +  {d \bar{y}^2}\big) + \frac{k}{\bar{u}^2} \big(d \bar{v}^2 +  {d \bar{u}^2}\big).\label{4.22}
\end{align}
which is precisely the original $AdS_2 \times H^2 \times H^2$ metric.
Consequently, the dual metric approaches the original $AdS_2\times H^2\times H^2$ metric
asymptotically, although the full dual geometry is not identical to the original one. Thus, the
original product geometry is recovered as the asymptotic metric of the non-Abelian T-dual background.
\\
\\
{\bf Interior region: $|\tilde{x}_2|<l^2$}
\\
We now consider the interior region, which is not covered by the hyperbolic parametrization in
equation \eqref{4.14}. In this region, $\Delta_{12}=l^4-{\tilde{x}_2}^2>0$, and a convenient coordinate transformation is
\begin{align}
{\tilde x_{_1}} = & l^2 t,    &{\tilde x_{_2}} = l^2 \cos \theta,~~~~~~~~~~~ ~~ \nonumber\\
{\tilde x_{_3}} = & k x,      &{\tilde x_{_4}} = k \sinh y,~~~~~~~~~~~~ \nonumber\\
{\tilde x_{_5}} = & k v,      &{\tilde x_{_6}} = k \sinh u.~~~~~~~~~~~~
\label{4.23}
\end{align}
Here, $\Delta_{12}=l^4 \sin^2 \theta$, so the transformation is well defined in the open interval $0<\theta < \pi$.
Substituting equation \eqref{4.23} into the dual metric \eqref{4.11}, we obtain
\begin{align}
d\tilde{s}^{2}= l^2\left(\frac{dt^2}{\sin^2\theta}-d\theta^2\right)\
+k\left(dy^2+\frac{dx^2}{\cosh^2y}\right)
+k\left(du^2+\frac{dv^2}{\cosh^2u}\right). \label{4.24}
\end{align}
The corresponding anti-symmetric tensor field is
\begin{align}
\tilde{B}=l^2\cot\theta ~dt\wedge d\theta
+k\tanh y ~dx\wedge dy +k\tanh u ~dv\wedge du. \label{4.25}
\end{align}
As in the exterior region, the dual three-form field strength vanishes, $\widetilde H=d\widetilde B=0$.
Thus, the interior parametrization does not introduce a non-zero three-form field strength.
The metric remains Lorentzian: the $t$-direction is spacelike, whereas the $\theta$-direction is timelike,
and the two hyperbolic factors contribute four additional spacelike directions.
Hence, the interior sector is not excluded by the signature of the dual metric.
Rather, it represents a distinct coordinate region in which the causal character of the two coordinates associated
with $(\tilde{x}_1,\tilde{x}_2)$ is interchanged relative to the exterior form \eqref{4.15}.

The metric  \eqref{4.24} is independent of $t, x$, and $v$, and therefore possesses the Killing vectors
$\partial_t, \partial_x$, and $\partial_v$. In this sector, $\partial_t$ is spacelike, with
$||\partial_t||^2= {\tilde G}_{tt}=l^2 /{\sin^2\theta}>0$,
while $\partial_x$ and $\partial_v$ are also spacelike.
The scalar curvature can be obtained from equation  \eqref{4.17} by replacing $\cosh r$ with $\cos \theta$ and $\sinh^2r$ with $-\sin^2\theta$.
It is given by
\begin{align}
\tilde {\cal R}=-\frac{1}{l^2}-\frac{1}{k}
+\frac{3+\cos^2\theta}{l^2\sin^2\theta}+\frac{5-3\sinh^2y}{2k\cosh^2y}
+\frac{5-3\sinh^2u}{2k\cosh^2u}. \label{4.26}
\end{align}
This expression is finite for $0<\theta <\pi$ and finite $y,u$, but diverges as $\theta \to 0$ or $\theta \to \pi$.
These limiting surfaces correspond respectively to $\tilde{x}_2=l^2$ and $\tilde{x}_2=-l^2$, where $\Delta_{12}$ vanishes.
The divergence of the curvature at both endpoints indicates that these boundaries are true curvature singularities,
rather than removable coordinate singularities. In contrast, the midpoint $\theta=\pi/2$, corresponding to $\tilde{x}_2=0$,
is regular with respect to the curvature invariant  \eqref{4.26}. Thus, the interior sector contains a regular open region bounded by two curvature-singular surfaces.
The interior geometry also exhibits a different causal structure from the exterior branch.
In particular, the Killing vector $\partial_t$, which is timelike in the exterior sector, becomes spacelike throughout the interior region.
The surfaces $\theta=0, \pi$ are singular boundaries, and no regular continuation through them is established by the coordinate transformation considered here. Consequently, the interior sector should be treated as a distinct Lorentzian region of the dual target space,
rather than being identified with the exterior geometry by a regular coordinate transformation.
Taken together, the exterior and interior parametrizations demonstrate that the dual background has a richer coordinate structure than is visible
from equation \eqref{4.14} alone.

\section{D-branes in the $AdS_2 \times H^2 \times H^2$ $\sigma$-model under the non-Abelian T-duality}
\label{Sec.V}

Before studying the worldsheet boundary conditions and $D$-brane configurations of the $AdS_2\times H^2\times H^2$ $\sigma$-model and its dual,
we briefly review the gluing matrix description of open string boundary conditions and its transformation under Poisson-Lie T-duality, following \cite{CC}.
We then discuss the role of the world-volume gauge field in the Neumann boundary condition and introduce the corresponding gauge-invariant two-form.
The detailed derivations and related geometric conditions are collected in three appendices.
Appendix \ref{app:A} explains the gauge invariance of the boundary coupling and derives the gauge-invariant two-form in terms of the pullback of the $B$-field and the world-volume field strength.
Appendix \ref{app:B} derives the Neumann boundary condition and the corresponding Neumann block of the gluing matrix from the open string action.
Finally, Appendix \ref{app.C} summarizes the geometric conditions on $D$-branes, including the Dirichlet and Neumann projectors and their integrability conditions in the Lie algebra frame.
\\
\\
{\bf Gluing matrix.} Consider a $d$-dimensional target space with the coordinates $X^\mu$ and background ${\cal E}_{\mu\nu}=G_{\mu\nu}+B_{\mu\nu}$.
We consider $Dp$-branes on this space with $d-(p +1)$
Dirichlet directions along which the fields $x^{i}$, $i = p+1, ..., d-1$, are frozen such that $x^i$ are the directions normal to the brane
at any given point on a $Dp$-brane.
In this way, the Dirichlet condition takes the following familiar form
\begin{eqnarray}
\partial_\tau x^i=0, \qquad \delta x^i=0, \qquad i=p+1,\ldots,d-1.  \label{5.1}
\end{eqnarray}
In addition, we choose local coordinates $x^{m}$, $m = 0,..., p$, as label Neumann directions which are coordinates on the brane.
Such a coordinate system is called adapted to the brane \cite{Zwiebach}.
The most general local boundary condition can be expressed in terms of a gluing matrix ${{\mathbb{R}}^{{\mu}}}_{{\nu}}$ as \cite{CC}:
\begin{equation}\label{5.2}
\partial_{-}X^{{\mu}} = {{\mathbb{R}}^{{\mu}}}_{{\nu}}\; \partial_{+} X^{{\nu}},
\end{equation}
where $\partial_{+} X^{{\mu}}$ and $\partial_{-} X^{{\mu}}$ are left- and right-moving fields, respectively, so that the
boundary relates them to each other.
The gluing matrix encodes the information about the Neumann and Dirichlet directions in its the eigenvalues and eigenvectors.
In coordinates adapted to the brane, it has the schematic block form
\begin{equation}\label{5.3}
\mathbb{R}\;=\;\left(
\begin{tabular}{c|c}
                 ${\mathbb R}_N$ & $0$ \\
\hline
                 $0$ & $-\mathbb{I}_D$ \\
                 \end{tabular} \right),
\end{equation}
where $\mathbb R_N$ is the Neumann-Neumann (NN) block and $-\mathbb I_D$ acts on the Dirichlet directions,
with $\mathbb{I}$ denoting the identity matrix.
Thus, the normal directions are $-1$ eigendirections of $\mathbb R$. In contrast, the tangent directions need not all be $+1$ eigendirections
of the full gluing matrix when the gauge-invariant two-form on the brane is non-zero.
In order to get more insight of the boundary conditions, one may write the gluing matrix in the Lie algebra frame.
Using the right-invariant one-forms, the boundary condition \eqref{5.2} becomes
\begin{align}\label{5.4}
R_{-}^a = {\mathbb{R}}^a_{~b} ~R_{+}^b,  \qquad {\mathbb{R}}^a_{~b} =R_\mu^{~a}~{\mathbb{R}}^\mu_{~\nu}~ (R^{-1})^{\nu}_{~b}.
\end{align}
This relation indicates that the object ${\mathbb{R}}^a_{~b}$ is a gluing map between currents at the worldsheet boundary.
Clearly, it is a map from the Lie algebra into itself. So, it may be assumed to be a constant Lie
algebra automorphism, i.e., it preserves the Lie algebra structure.
The projectors $\mathbb Q$ and $\mathbb N$ are transformed to the same frame in the analogous way; their geometric properties are summarized in
Appendix \ref{app.C}.

On the other hand, the non-zero NN block is constrained by the projected Neumann condition in the Lie algebra frame.
In the notation of the original construction, this gives ${\mathbb{R}}_{N}=(\mathbb{N}^T E \mathbb{N})^{-1}~(\mathbb{N}^T E \mathbb{N})^T$, where
$\mathbb N$ is the Neumann projector and $E$ is the bilinear form introduced in the construction of the $\sigma$-model (Eq. \eqref{3.8}).
Hence, the gluing matrix takes the block form \cite{CC}
\begin{eqnarray}\label{5.5}
{\mathbb{R}}\;=\;\left(
\begin{tabular}{c|c}
                 $(\mathbb{N}^T E \mathbb{N})^{-1}~(\mathbb{N}^T E \mathbb{N})^T$ & $0$ \\
\hline
                 $0$ & $-\mathbb{I}_D$ \\
                 \end{tabular} \right),
\end{eqnarray}
where the superscript ``T'' means transposition of the matrix.
For a space-filling brane, where all directions are the Neumann, $\mathbb N=\mathbb I$ and $\mathbb Q=0$,
the gluing matrix is, schematically, given by
\begin{eqnarray}\label{5.6}
{\mathbb{R}} = {E}^{-1} ~ {E}^{T}.
\end{eqnarray}
The relation above is the background matrix form of the Neumann gluing condition.
Below we make explicit how it is generalized when the independent world-volume field strength is retained.

\medskip
\noindent\textbf{Duality transformation of the gluing matrix.}
The Poisson-Lie transformation of the boundary condition follows by applying the canonical transformation
between the original and dual phase space variables \cite{sfetsos,sfetsos1}.
The canonical transformation between the pairs of variables $(R_\sigma^a~,~P_a)$ and $\big(({\tilde R}_\sigma)_a~,~{\tilde P}^a)$ is given by \cite{sfetsos1}
\begin{eqnarray}
R_\sigma^a &=& (\delta^a_{~b} - \Pi^{ac} {\tilde \Pi}_{cb}){\tilde P}^b- \Pi^{ab} ({\tilde R}_\sigma)_b,\label{5.7}\\
P_a &=& {\tilde \Pi}_{ab} {\tilde P}^b + ({\tilde R}_\sigma)_a,\label{5.8}
\end{eqnarray}
where $P_a=(R^{-1})^\mu{}_aP_\mu$ and $\widetilde P^a=(\widetilde R^{-1})^{\mu a}\widetilde P_\mu$ are the momenta in the invariant frames. Here $P_\mu$ and $\widetilde P_\mu$ are conjugate to $X^\mu$ and $\widetilde X^\mu$, respectively. Introducing light-cone components,
\begin{align}
R_\sigma^a =& \frac{1}{2}  (R_+^a - R_-^a), ~~~ &({\tilde R}_\sigma)_a = \frac{1}{2} ({\tilde R}_+{_a} - {\tilde R}_-{_a} ),~~~~~\label{5.9}\\
P_a =& \frac{1}{2} (E_{ba} R_+^b + E_{ab} R_-^b), ~~~~~~~~&{\tilde P}^a= \frac{1}{2} ({\tilde E}^{ba} {\tilde R}_+{_b}
+ {\tilde E}^{ab} {\tilde R}_-{_b}),\label{5.10}
\end{align}
one obtains, after substituting into the second canonical relation and using the background identities \eqref{3.8} and \eqref{3.12},
\begin{align}
E_{ba} R_+^b  =& ({\delta_a}^c  + {\tilde \Pi}_{ab} {\tilde E}^{cb}) {\tilde R}_+{_c},\label{5.11}\\
E_{ab} R_-^b  =& - ({\delta_a}^c  - {\tilde \Pi}_{ab} {\tilde E}^{bc}) {\tilde R}_-{_c}. \label{5.12}
\end{align}
The use of the Poisson-Lie background relation \eqref{3.12} then gives
\begin{align}
{\tilde R}_+{_a}  =& ({{{\tilde E}}^{{-1}})_{ba}}~ ({E_0}^{-1})^{cb}~E_{dc}~ R_+^d,\label{5.13}\\
{\tilde R}_-{_a}  =& - ({{{\tilde E}}^{{-1}})_{ab}}~ ({E_0}^{-1})^{bc}~E_{cd}~ R_-^d. \label{5.14}
\end{align}
Substituting the original gluing relation $R_-^a=\mathbb R^a{}_bR_+^b$ into Eq.~\eqref{5.14}, and then using Eq.~\eqref{5.13},
yields the general Poisson-Lie transformation law for the gluing matrix \cite{CC}:
\begin{eqnarray}\label{5.15}
{\tilde {\mathbb{R}}}_a^{~b} = - ({{{\tilde E}}^{{-1}})_{ac}}~ ({E_0}^{-1})^{cd}~E_{de}~ {\mathbb{R}}^e_{~f}~
({E}^{-1})^{hf}~{E_0}_{gh}~{\tilde E}{^{bg}},
\end{eqnarray}
where ${\tilde {\mathbb{R}}}_a^{~b}$ denotes the dual gluing matrix.
In particular, $\det \widetilde{\mathbb R}=\det(-\mathbb R)$, which gives a preliminary restriction on the possible dual brane dimensions.
It does not, however, determine the multiplicity of the eigenvalue $-1$ and hence cannot replace the exact spectral and projector analysis.

For the semi-Abelian double used in the present example, corresponding to non-Abelian T-duality, one has $E(g)=E_0=E_0^T$. The NN block in Eq.~\eqref{5.5} therefore reduces to $\mathbb R_N=\mathbb I_N$, and the original gluing matrix becomes
\begin{eqnarray}\label{5.16}
{\mathbb{R}}=\left(
\begin{tabular}{c|c}
                 $\mathbb{I}$ & $0$ \\
\hline
                 $0$ & $-\mathbb{I}$ \\
                 \end{tabular} \right).
\end{eqnarray}
The general Poisson-Lie  transformation law then simplifies to
\begin{equation}
\widetilde{\mathbb R}=-\widetilde E^{-1}\mathbb R\widetilde E^T, \label{5.17}
\end{equation}
where $\widetilde E$ is the dual background matrix given earlier in the paper.

\medskip
\noindent\textbf{Open string boundary coupling and the gauge-invariant two-form.}
The gluing relation must be supplemented by the coupling of the open string endpoints to the $U(1)$ gauge potential on the $D$-brane.
Let $\iota: \cal W \to \M$ be the embedding of the world-volume and let
$x: \partial\Sigma \to \cal W$ denotes the boundary map, so that $X|_{\partial\Sigma}=\iota \circ x$.
We write the world-volume potential and its field strength as \cite{Johnson}
\begin{equation}
A_{\mathrm{wv}}=A_m(x)\,\mathrm dx^m, \qquad F_{\mathrm{wv}}=\mathrm dA_{\mathrm{wv}}. \label{5.18.1}
\end{equation}
The endpoint coupling is included in the open string action together with the bulk metric and $B$-field terms.
With the normalization used here, the factor $2\pi\alpha'$ is absorbed into $A_{\mathrm{wv}}$.
Under $B \mapsto B+\mathrm d \mathbf\Lambda$, the boundary variation of the bulk $B$-term is cancelled
by $A_{\mathrm{wv}}\mapsto A_{\mathrm{wv}}-\iota^\ast \mathbf\Lambda$.
Hence the combination entering the boundary dynamics is the gauge-invariant two-form \cite{Johnson}
\begin{equation}
{\mathcal F=\iota^*B+F_{\mathrm{wv}}.} \label{5.19.1}
\end{equation}
The explicit component form of the action, the pullback notation, and the gauge-invariance argument leading
to Eq.~\eqref{5.19.1} are given in Appendix \ref{app:A}.

The appearance of $\mathcal F$ in the Neumann condition follows directly from the first variation of the open string action.
In adapted coordinates, the allowed boundary variations vanish in the Dirichlet directions, $\delta x^i=0$, but remain arbitrary along the brane, $\delta x^m\neq0$. The metric part of the bulk action contributes a term proportional to $G_N \partial_\sigma x^m$, whereas the variation of the bulk $B$-field term contributes a term proportional to $-(\iota^\ast B)\partial_\tau x^m$.
The variation of the endpoint coupling to $A_{\mathrm{wv}}$ supplies the additional contribution $-F_{\mathrm{wv}}\partial_\tau x^m$.
Thus, the three boundary contributions combine into the single expression
\begin{equation*}
\left.\delta S_{\mathrm{open}}\right|_{\partial\Sigma,N}
=\frac{1}{2\pi\alpha'}\int_{\partial\Sigma}\mathrm d\tau\,\delta x^m
\bigg[(G_N)_{mn}\partial_\sigma x^n-\mathcal F_{mn}\partial_\tau x^n\bigg].
\end{equation*}
Since the Neumann variations are arbitrary, the coefficient of each $\delta x^m$ must vanish. This gives the mixed Neumann condition in $(\tau,\sigma)$ coordinates. Rewriting it in terms of the light-cone derivatives $\partial_\pm=\partial_\tau\pm\partial_\sigma$ gives \cite{Stanciu33,Cheung}
\begin{equation}
(G_N+\mathcal F)\partial_-x^m=(G_N-\mathcal F)\partial_+x^m, \label{5.20.1}
\end{equation}
where $G_N$ is the metric induced on the tangent subspace of the brane.
Comparing this equation with $\partial_-x^m=\mathbb R_N\partial_+x^m$ yields the Neumann block \cite{Stanciu33,Cheung}
\begin{equation}
\mathbb R_N=(G_N+\mathcal F)^{-1}(G_N-\mathcal F), \label{5.21.1}
\end{equation}
where the inverse is taken on the Neumann subspace. In particular, $\mathbb R_N=\mathbb I_N$ when $\mathcal F=0$,
whereas a non-zero $\mathcal F$ generally makes the Neumann block non-trivial.
Detailed derivations of Eqs.~\eqref{5.20.1} and~\eqref{5.21.1}, including the variations of the metric, $B$-field,
and endpoint terms, as well as the conversion to light-cone coordinates, are provided in Appendix \ref{app:B}.
The relation between Eqs.~\eqref{5.5} and~\eqref{5.21.1} is worth making explicit. In a common frame, when no independent world-volume field strength is present, $\mathcal F=\iota^*B$. Since the metric is symmetric and the $B$-field is antisymmetric, the projected background matrix and its transpose are then $G_N+\iota^*B$ and $G_N-\iota^*B$, respectively. Thus, Eq.~\eqref{5.5} is the projected background matrix expression of Eq.~\eqref{5.21.1} in the case $F_{\mathrm{wv}}=0$.
When $F_{\mathrm{wv}}$ is retained explicitly, the two-form entering the Neumann block is the full $\mathcal F$, not merely $\iota^*B$.
All matrices must first be expressed in the same frame and projected onto the same Neumann distribution before the two descriptions are compared.

\medskip
\noindent\textbf{Application to the dual background.}
For the dual brane, the corresponding gauge-invariant two-form is
\begin{equation}
\widetilde{\mathcal F}=\iota^*\widetilde B+\widetilde F_{\mathrm{wv}}, \label{5.22.1}
\end{equation}
and its Neumann gluing block satisfies
\begin{equation}
\widetilde{\mathbb R}_N =(\widetilde G_N+\widetilde{\mathcal F})^{-1} (\widetilde G_N-\widetilde{\mathcal F}). \label{5.23.1}
\end{equation}
Conversely, whenever $\mathbb I_N+\widetilde{\mathbb R}_N$ is invertible, the two-form can be reconstructed as
\begin{equation}
\widetilde{\mathcal F} =\widetilde G_N(\mathbb I_N-\widetilde{\mathbb R}_N)
(\mathbb I_N+\widetilde{\mathbb R}_N)^{-1}. \label{5.24.1}
\end{equation}
The inverse is on the Neumann subspace and the order of matrix factors is important. The independent world-volume field strength is then $\widetilde F_{\mathrm{wv}}=\widetilde{\mathcal F}-\iota^*\widetilde B$. Locally, it must be closed, $\mathrm d\widetilde F_{\mathrm{wv}}=0$, since it is the field strength of a potential.
More generally, $\mathrm d\widetilde{\mathcal F}=\iota^*\widetilde H$, where $\widetilde H=\mathrm d\widetilde B$.
For the dual background considered here, $\widetilde H=0$, so $\widetilde{\mathcal F}$ is closed as well.
Notice that $\widetilde F_{\mathrm{wv}}$ need not be non-zero in every branch:
it vanishes whenever the reconstructed $\widetilde{\mathcal F}$ equals the pullback of $\widetilde B$.
In using Eqs.~\eqref{5.23.1} and~\eqref{5.24.1},
the Neumann block of $\widetilde{\mathbb R}$, the induced metric $\widetilde G_N$,
and the pullback $\iota^*\widetilde B$ must be expressed in one and the same frame.
In particular, the gluing matrix obtained from the Poisson-Lie transformation is initially written in the Lie algebra frame;
it must be projected onto the dual Neumann distribution and converted consistently before the world-volume two-form is reconstructed.

\subsection{D-branes in the $AdS_2 \times H^2 \times H^2$ $\sigma$-model and its dual pair}

We now apply the resulting dual gluing matrix to the $AdS_2\times H^2\times H^2$ $\sigma$-model.
The original and dual models are constructed on the semi-Abelian double $(\A_2\oplus\A_2\oplus\A_2,6\A_1)$,
with the commutation relations and background fields given earlier.
The analysis of each candidate dual brane therefore proceeds by determining the exact spectrum of $\widetilde{\mathbb R}$,
constructing the complementary Dirichlet and Neumann projectors, and checking the projector, metric orthogonality,
and Neumann integrability conditions summarized in Appendix \ref{app.C}. The spectrum identifies candidate Dirichlet directions
through the multiplicity of the eigenvalue $-1$, but does not by itself establish a consistent boundary configuration.
One must also verify the projected Neumann condition and use Eq.~\eqref{5.24.1} to reconstruct $\widetilde{\mathcal F}$,
after which $\iota^*\widetilde B$ is subtracted to determine $\widetilde F_{\mathrm{wv}}$.
This combined algebraic, geometric,
and gauge field analysis is the basis for the seven classes of $D$-brane configurations considered below.

\medskip
\noindent\textbf{Case 1: A $D{(-1)}$-brane with Dirichlet directions in all six coordinate directions.}
This case corresponds to a $D{(-1)}$-brane on the original manifold, for which all six directions are Dirichlet.
The original gluing matrix is $ {\mathbb{R}}= - \mathbb{I}_{6 \times 6}$, and its Neumann and Dirichlet projectors are
${\mathbb{N}} = 0$ and ${\mathbb{Q}} = \mathbb{I}_{6 \times 6}$, respectively.
Using the original gluing condition \eqref{5.4} together with the right-invariant one-forms of the original group manifold,
one obtains
\begin{eqnarray*}
\partial_\tau x_i=0,\qquad
e^{x_i}\partial_\tau x_j=0,
\qquad i=1,3,5,\quad j=i+1.
\end{eqnarray*}
Since $e^{x_i}\neq 0$, these equations are equivalent to $\partial_\tau x_a=0$ for all $a=1,\ldots,6$,
as expected for a pointlike $D(-1)$-brane.
Using the duality transformation \eqref{5.17}, we obtain the dual gluing matrix
\begin{eqnarray}\label{5.25.0}
{\tilde {\mathbb{R}}} = {{\tilde E}}^{{-1}}~  {{{\tilde E}}^{{T}}},
\end{eqnarray}
which is the dual counterpart of \eqref{5.6}. Hence,
${\tilde {\mathbb{R}}}$ describes a space-filling brane on ${\tilde G}$,
with Neumann boundary conditions in all directions.
This brane is dual to a pointlike brane on the original group manifold; accordingly, the $D{(-1)}$-brane is mapped to a $D{5}$-brane.
The dual gluing matrix has a block-diagonal form
\begin{eqnarray}\label{5.26.0}
{\widetilde {\mathbb{R}}} = {\widetilde {\mathbb{R}}}_{12} \oplus {\widetilde {\mathbb S}}_{34} \oplus {\widetilde {\mathbb S}}_{56},
\end{eqnarray}
where
\begin{eqnarray}\label{5.27.0}
{\widetilde {\mathbb{R}}}_{12}
=\frac{1}{\Delta_{_{12}}}\left( \begin{array}{cc}
                   {l^4 + {\tilde x}_{_2}^2} & {2 l^2 {\tilde x}_{_2}} \\
                    {2 l^2{\tilde x}_{_2}} & {l^4 + {\tilde x}_{_2}^2}
                      \end{array} \right),
\end{eqnarray}
and
\begin{eqnarray}\label{5.28.0}
{\widetilde {\mathbb S}}_{34} = \frac{1}{{\Delta_{34}}}
\left( \begin{array}{cc}
{k^2-\widetilde x_4^2} & -{2k\widetilde x_4}\\
{2k\widetilde x_4}& {k^2-\widetilde x_4^2}
\end{array} \right),
\qquad
{\widetilde {\mathbb S}}_{56}= \frac{1}{{\Delta_{56}}}
\left( \begin{array}{cc}
{k^2-\widetilde x_6^2} & -{2k\widetilde x_6}\\
{2k\widetilde x_6}& {k^2-\widetilde x_6^2}
\end{array} \right),
\end{eqnarray}
with $\Delta_{12}$, $\Delta_{34}$, and $\Delta_{56}$ defined in Section \ref{Sec.IV}.
We restrict the discussion to the regular coordinate region $\Delta_{12} \neq 0$.
Note that the determinant of ${\widetilde {\mathbb{R}}}$ is equal to 1.
We first determine the exact eigenvalues of the dual gluing matrix. The eigenvalues of the ${\widetilde {\mathbb{R}}}_{12}$ sub-matrix are
\begin{eqnarray}\label{5.29.0}
\lambda_{12}^{\pm} =\frac{l^2 \pm \widetilde x_2}{l^2\mp\widetilde x_2}.
\end{eqnarray}
The corresponding eigenvectors in the complete six-dimensional space are
$V_{12}^{+}=\operatorname{diag}(1 , 1 , 0 , 0 , \\0 , 0)^T$, and $V_{12}^{-}=\operatorname{diag}(1 , -1 , 0 , 0 , 0 , 0)^T$, respectively.
The other two blocks have eigenvalues
\begin{eqnarray}\label{5.30.0}
\lambda_{4}^{\pm} = \frac{k^2 - {\widetilde x_4}^2 \pm 2i k  {\widetilde x_4}}{k^2 + {\widetilde x_4}^2}= e^{\pm 2i\arctan(\frac{\widetilde x_4}{k})},
\end{eqnarray}
and
\begin{eqnarray}\label{5.31.0}
\lambda_{6}^{\pm} = \frac{k^2 - {\widetilde x_6}^2 \pm 2i k  {\widetilde x_6}}{k^2 + {\widetilde x_6}^2}=e^{\pm 2i\arctan(\frac{\widetilde x_6}{k})}.
\end{eqnarray}
The corresponding eigenvectors in the full coordinate space may be chosen as
$V_{4}^{\pm}=\operatorname{diag} (0 , 0 , 1 ,\\ \mp i , 0 , 0)^T$ and $V_{6}^{\pm}=\operatorname{diag}(0 , 0 , 0 , 0 , 1 , \mp i)^T$, respectively.
For finite real values of the dual coordinates in the regular region, none of these six eigenvalues equals $-1$.
Consequently, the dual gluing matrix possesses no Dirichlet-type eigenvalues, confirming that the corresponding dual brane
is a space-filling $D5$-brane, as anticipated earlier.
This conclusion follows from the exact spectrum, rather than the determinant alone.
We next verify the dual coordinate boundary conditions. Since the dual group is Abelian, its right-invariant one-forms reduce to
${\tilde R}_{{\pm}_a}=\partial_{_\pm} {\tilde x}_{_{a}}$. Substituting the dual gluing matrix from equations \eqref{5.26.0}-\eqref{5.28.0}
into the dual gluing condition, ${\tilde R}_-{_a}= {\tilde {\mathbb{R}}}_a^{~b} {\tilde R}_+{_b}$,
gives the following coordinate boundary conditions
\begin{align}
l^4 \partial_{\sigma} {\tilde x}_{_i}+{\tilde x}_{_2}^2 \partial_{\tau} {\tilde x}_{_i} +l^2 {\tilde x}_{_2}(\partial_{\tau} {\tilde x}_{_j} +
\partial_{\sigma} {\tilde x}_{_j}) =&0,~~~~i,j=1,2;~i \neq j,\nonumber\\
k^2 \partial_{\sigma} {\tilde x}_{_l}-{\tilde x}_{_4}^2 \partial_{\tau} {\tilde x}_{_l} -(-1)^{^m} k  {\tilde x}_{_4}(\partial_{\tau} {\tilde x}_{_m} +
\partial_{\sigma} {\tilde x}_{_m}) =&0,~~~~l,m=3,4;~l \neq m,\nonumber\\
k^2 \partial_{\sigma} {\tilde x}_{_n}-{\tilde x}_{_6}^2 \partial_{\tau} {\tilde x}_{_n} -(-1)^{^p}  k  {\tilde x}_{_6}(\partial_{\tau} {\tilde x}_{_p} +
\partial_{\sigma} {\tilde x}_{_p}) =&0,~~~~n,p=5,6;~n \neq p.\nonumber
\end{align}

The dual Dirichlet and Neumann projectors are therefore
\begin{eqnarray}\label{5.32.0}
\widetilde {\mathbb Q}=0,\qquad \widetilde {\mathbb N}={\mathbb{I}}_{_{6 \times 6}}.
\end{eqnarray}
Condition \eqref{brane13.2} is satisfied trivially because $\widetilde {\mathbb Q}=0$.
The Neumann condition \eqref{brane13.3}, with $\widetilde {\mathbb N}={\mathbb{I}}_{_{6 \times 6}}$, reduces to the matrix equation
${\widetilde E}^T - {\widetilde E} {\widetilde {\mathbb R}}=0 $, which coincides with \eqref{5.25.0}.
Thus, the Neumann condition \eqref{brane13.3} is satisfied identically wherever $\widetilde E$ is invertible.
The orthogonality condition \eqref{brane13.4} is also satisfied, since ${{\widetilde {\mathbb N}^\mu}_{{{\rho}}}}\;
{\widetilde G}_{\mu \nu}\;{{\widetilde {\mathbb Q}}^{\nu}}_{{\;\sigma}}= 0$.
The Neumann integrability condition \eqref{brane13.5} is satisfied because all components of $\widetilde {\mathbb N}$
are constant in the dual coordinates and therefore have vanishing derivatives.
Equivalently, the Neumann distribution spans the full tangent space of the dual target manifold and is integrable.

To confirm the Neumann boundary conditions, we analyze the gauge-invariant two-form on the dual $D5$-brane from equations \eqref{5.22.1} and \eqref{5.24.1}.
In this way, we use the exact dual gluing matrix obtained above.
Since the dual brane is space-filling, all six target space directions are Neumann directions.
Consequently, the Neumann block of the gluing matrix coincides with the full dual gluing matrix, namely, ${\widetilde {\mathbb R}}_N ={\widetilde {\mathbb R}}$.
Therefore, the gauge-invariant two-form can be reconstructed from the Neumann gluing relation \eqref{5.24.1}.
Substituting the exact dual metric and dual gluing matrix into this expression, and evaluating the three independent coordinate blocks,
we obtain
\begin{align}\label{5.33.0}
\widetilde{\mathcal F}= -\frac{\widetilde x_2}{\Delta_{12}}
d\widetilde x_1\wedge d\widetilde x_2 +\frac{\widetilde x_4}{\Delta_{34}}
d\widetilde x_3\wedge d\widetilde x_4+\frac{\widetilde x_6}{\Delta_{56}} d\widetilde x_5\wedge d\widetilde x_6,
\end{align}
Since the dual brane fills the entire target space, the embedding map is the identity,
and hence the pullback of the $\widetilde B$-field is simply $\iota^\ast \widetilde B=\widetilde B$.
Using the definition \eqref{5.22.1} and comparing the resulting expression for $\widetilde{\mathcal F}$
with the $\widetilde B$-field given in equation \eqref{4.12}, we find that the two forms coincide. It therefore follows that
${{\widetilde F}_{\mathrm{wv}}=0}$.
Thus, although the gauge-invariant two-form on the dual $D5$-brane is generally non-zero,
no additional world-volume gauge field strength is required in this case.
The non-trivial Neumann gluing matrix is reproduced entirely by the pullback of the $\widetilde B$-field.

Finally, since the field strength $\widetilde H=d\widetilde B$ vanishes, the gauge-invariant two-form is closed:
$d {\mathcal  {\widetilde F}} = \iota^\ast  \widetilde{H} =0$.
We conclude that Case 1 describes the duality between a pointlike $D(-1)$-brane in the original model and
a space-filling $D5$-brane in the dual model. Although the dual $D5$-brane has a non-zero gauge-invariant two-form,
no independent world-volume gauge field strength is required in the present construction.
The dual configuration therefore satisfies the projector, boundary condition, and integrability criteria considered in Appendix \ref{app.C}.

\medskip
\noindent\textbf{Case 2: A $D0$-brane with one Neumann and five Dirichlet directions.}
Case 2 corresponds to an original $D0$-brane with one Neumann direction and five Dirichlet directions.
The original gluing matrix is
\begin{align}\label{5.34.2}
\mathbb R=\operatorname{diag}(1,-1,-1,-1,-1,-1).
\end{align}
The corresponding Dirichlet and Neumann projectors are given respectively by
\begin{align}\label{5.35.2}
\mathbb{Q} &= \operatorname{diag}(0, 1, 1, 1, 1, 1),\nonumber\\
\mathbb{N} &= \operatorname{diag}(1, 0, 0, 0, 0, 0),
\end{align}
Using the right-invariant one-forms given in \eqref{4.3}, together with the original gluing condition \eqref{5.4},
we obtain the boundary conditions
\begin{align}
\partial_\sigma x_1 &=0, \qquad ~~ \partial_\tau x_3 =0, \qquad \qquad  \partial_\tau x_5=0,\nonumber\\
e^{x_1} \partial_\tau x_2&=0, \qquad e^{x_3}\partial_\tau x_4=0,  \qquad e^{x_5}\partial_\tau x_6=0.\label{5.36.2}
\end{align}
Since the exponential factors are non-zero, these conditions describe a $D0$-brane whose world-volume lies along the $x_1$ direction.
Under the non-Abelian T-duality transformation, the dual gluing matrix is
\begin{align}\label{5.37.2}
\widetilde{\mathbb R}
=-\widetilde{E}^{-1} {\mathbb R} \widetilde{E}^{T} = \operatorname{diag}(-1,1) \oplus {\widetilde {\mathbb S}}_{34} \oplus {\widetilde {\mathbb S}}_{56},
\end{align}
where ${\widetilde {\mathbb S}}_{34}$ and $ {\widetilde {\mathbb S}}_{56}$ are given by \eqref{5.28.0}.
Since the determinant of the dual gluing matrix is $\det \widetilde{\mathbb{R}} = -1$,
the dual brane is expected to possess one, three, or five Dirichlet directions, corresponding to a $D4$-, $D2$-, or $D0$-brane, respectively.
We note that the determinant of the dual gluing matrix is not sufficient to determine the $D$-brane dimension.
Although the determinant provides a useful consistency check, it does not determine the multiplicity of the eigenvalue $-1$,
the corresponding Dirichlet subspace, the Neumann projector, or the integrability of the resulting distribution.
The original $D0$-brane lies along one of the coordinate directions (one Neumann direction) in the original manifold,
whereas the dual branes are non-trivially embedded in the dual manifold;
this embedding can be obtained explicitly by determining the complete spectrum of the eigenvalues of ${\widetilde {\mathbb{R}}}$.
The complete spectrum of the eigenvalues in this case is
\begin{eqnarray}\label{5.38.2}
\textrm{spec}({\mathbb R})=\Big\{-1 ,1,  e^{\pm 2i\arctan(\frac{\widetilde x_4}{k})}, e^{\pm 2i\arctan(\frac{\widetilde x_6}{k})}\Big\}.
\end{eqnarray}
For finite real ${\widetilde x}_{4}$, ${\widetilde x}_{6}$ and $k > 0$, neither of the last two blocks has an eigenvalue equal to $-1$.
They therefore do not represent additional Dirichlet directions.
Thus, the dual gluing matrix has exactly one Dirichlet eigenvalue, irrespective of the values of these two coordinates.
It follows that the generic dual brane has five Neumann directions and one Dirichlet direction.
In six target space dimensions, its correct interpretation is therefore a $D4$-brane, not the alternatives
$D0$-, and $D2$-branes previously inferred from the determinant alone.
Thus, the dual brane is a $D4$-brane. Its embedding is specified locally by
\begin{eqnarray}\label{5.39.2}
\widetilde x_1=c,
\end{eqnarray}
where $c$ is a constant. The exact eigenvector with eigenvalue $-1$ is the coordinate vector in the $\widetilde x_1$ direction.
We furthermore define the corresponding projectors as
\begin{eqnarray}\label{5.40.2}
\widetilde{\mathbb Q}=\operatorname{diag}(1,0,0,0,0,0), \qquad \widetilde{\mathbb N} =\operatorname{diag}(0,1,1,1,1,1).
\end{eqnarray}
These matrices satisfy the projector identities ${\widetilde{\mathbb Q}}^2=\widetilde{\mathbb Q}, {\widetilde{\mathbb N}}^2=\widetilde{\mathbb N}$, and
${\widetilde{\mathbb Q}} {\widetilde{\mathbb N}}={\widetilde{\mathbb N}} \widetilde{\mathbb Q}=0$. Furthermore, direct multiplication gives
${\widetilde{\mathbb R}} {\widetilde{\mathbb Q}}=  {\widetilde{\mathbb Q}} {\widetilde{\mathbb R}}=- {\widetilde{\mathbb Q}}$, as required for the Dirichlet projector.
We now verify that the Neumann condition \eqref{brane13.3} is satisfied.
Since the first dual direction is Dirichlet, the Neumann part consists of the five directions $(\widetilde x_2,\widetilde x_3,\widetilde x_4,\widetilde x_5,\widetilde x_6)$. The projection of the dual background matrix onto this part is
\begin{align}\label{5.41.2}
\big(-\frac{l^2}{\Delta_{12}}\big) \oplus {\widetilde E}^{(34)} \oplus {\widetilde E}^{(56)},
\end{align}
where ${\widetilde E}^{(34)}$ and $ {\widetilde E}^{(56)}$ are given by the second equation of \eqref{4.10.1}.
The Neumann block of the gluing matrix is
\begin{align}\label{5.43.2}
\widetilde{\mathbb R}_N =1 \oplus {\widetilde {\mathbb S}}_{34} \oplus {\widetilde {\mathbb S}}_{56}.
\end{align}
Direct matrix multiplication gives ${\widetilde E}^{(34)\,-1}\,{\widetilde E}^{(34)\,{T}}={\widetilde {\mathbb S}}_{34}$, and
${\widetilde E}^{(56)\,-1}\,{\widetilde E}^{(56)\,{T}}={\widetilde {\mathbb S}}_{56}$,  while the one-dimensional $\widetilde x_2$ sector satisfies
$\big(-{l^2}/{\Delta_{12}}\big)^{-1} \big(-{l^2}/{\Delta_{12}}\big)=1$.
Therefore, $\widetilde{\mathbb R}_N =\widetilde{\mathbb E}_N^{-1} \widetilde{\mathbb E}_N^{T}$, or, equivalently,
\begin{align}\label{5.44.2}
\widetilde{\mathbb E}_N^{T} -\widetilde{\mathbb E}_N \widetilde{\mathbb R}_N=0.
\end{align}
This is precisely the Neumann condition \eqref{brane13.3} restricted to the dual $D4$-brane.
Hence, the Neumann condition is satisfied on the regular patch where the relevant matrices are invertible.
As noted above, the dual brane is locally embedded as $\widetilde x_1=c$.
Its tangent distribution is generated by $\{\partial/\partial {\widetilde x_2} , \partial/\partial {\widetilde x_3} , \partial/\partial {\widetilde x_4} , \partial/\partial {\widetilde x_5} , \partial/\partial {\widetilde x_6}\}$.
This distribution is integrable. Indeed, the coordinate vector fields commute, and the Lie bracket of any two vector fields tangent to $\widetilde x_1=c$
remains tangent to the same hypersurface. Equivalently, the Neumann distribution satisfies the integrability condition stated in Appendix \ref{app.C}.
The metric-orthogonality condition is also satisfied. Since the exact dual metric \eqref{4.11} is diagonal in these coordinates,
it has no mixed metric components between the $\widetilde x_1$ direction and the remaining five directions. Hence,
${{\widetilde {\mathbb N}^\mu}_{{{\rho}}}}\;
{\widetilde G}_{\mu \nu}\;{{\widetilde {\mathbb Q}}^{\nu}}_{{\;\sigma}}= 0$.
Therefore, the $D4$-brane interpretation in Case 2 is supported by an explicit Dirichlet projector,
a complementary Neumann projector, and an integrable tangent distribution, rather than by the determinant alone.

Finally, we determine the gauge-invariant two-form on the dual world-volume, which follows from the Neumann gluing matrix defined in \eqref{5.43.2}
and formula \eqref{5.24.1}.
The $\widetilde x_2$ contribution vanishes because its Neumann gluing eigenvalue is $+1$. Evaluating the ${\widetilde {\mathbb S}}_{34}$ and
${\widetilde {\mathbb S}}_{56}$ blocks gives
\begin{align}\label{5.45.2}
\widetilde{\mathcal F}=\frac{\widetilde x_4}{\Delta_{34}} d\widetilde x_3\wedge d\widetilde x_4
+\frac{\widetilde x_6}{\Delta_{56}} d\widetilde x_5\wedge d\widetilde x_6.
\end{align}
On the $D4$-brane embedding $\widetilde x_1=c$, we have $d\widetilde x_1=0$. Therefore, the pullback of the $\widetilde B$-field is
$\iota^\ast \widetilde{B}  =\widetilde{\mathcal F}$.
Using the definition \eqref{5.22.1}, we finally obtain ${{\widetilde F}_{\mathrm{wv}}=0}$.
Thus, for Case 2, the original $D0$-brane is mapped to a $D4$-brane embedded at $\widetilde x_1=c$.
The projected Neumann condition \eqref{brane13.3} is satisfied, and the gauge-invariant world-volume two-form coincides with
the pullback of the $\widetilde B$-field, without requiring an additional world-volume gauge field strength.

\medskip
\noindent\textbf{Case 3: A $D1$-brane with two Neumann and four Dirichlet directions.}
In this case, the original gluing matrix is ${\mathbb{R}}=$diag$({\mathbb{I}}_{_{2 \times 2}} , -{\mathbb{I}}_{_{4 \times 4}})$,
corresponding to a $D1$-brane with two Neumann and four Dirichlet directions.
Applying the duality transformation \eqref{5.17}, we obtain the dual gluing matrix, giving
\begin{eqnarray}\label{5.46}
{\widetilde {\mathbb{R}}} = (-{\widetilde {\mathbb{R}}}_{12}) \oplus {\widetilde {\mathbb S}}_{34} \oplus {\widetilde {\mathbb S}}_{56},
\end{eqnarray}
where ${\widetilde {\mathbb{R}}}_{12}$ is given by \eqref{5.27.0},
while ${\widetilde {\mathbb S}}_{34}$ and $ {\widetilde {\mathbb S}}_{56}$ are given by \eqref{5.28.0}.
The determinant is $\det {\tilde {\mathbb{R}}} = 1$, so the the dual brane may include the following directions:
\\
(1) Six Dirichlet directions, i.e., the dual brane is a $D(-1)$-brane.
In this case, ${\tilde {\mathbb{R}}}$ has six $-1$ eigenvalues. This
means it can be diagonalized to take the form ${\tilde {\mathbb{R}}} =- \mathbb{I}$.
\\
(2) Four Dirichlet directions and two Neumann directions. This is a $D1$-brane, so the $D1$-brane can be dual to a $D1$-brane.
\\
(3) Two Dirichlet directions and four Neumann directions, i.e., the dual brane can be a $D3$-brane.
\\
(4) Zero Dirichlet directions. This is a $D5$-brane whose embedding in ${\tilde G}$ is given by ${\tilde {\mathbb{R}}}$.

The above list provides only preliminary possibilities inferred from the determinant of the dual gluing matrix.
However, the determinant alone is not sufficient to establish which of these $D$-brane configurations are actually realized. In
particular, the determinant gives only the product of the eigenvalues and does not determine the multiplicity of the eigenvalue $-1$,
which identifies the Dirichlet directions. Moreover, in the presence of a non-zero gauge-invariant world-volume two-form,
the Neumann block of the gluing matrix need not have eigenvalues restricted to $\pm 1$.
Consequently, a more detailed analysis requires calculating the exact eigenvalues and their corresponding eigenvectors.
To determine the actual $D$-brane configurations in this case, we therefore examine the spectrum of the dual gluing matrix.
The first block of $\widetilde{\mathbb R}$ has eigenvalues
\begin{eqnarray}\label{5.48}
\lambda_{12}^{\pm} =-\frac{l^2\pm\widetilde x_2}{l^2\mp\widetilde x_2}.
\end{eqnarray}
In six dimensions, the corresponding eigenvectors can be written as $V_{12}^{\pm} =\operatorname{diag}(1 , \pm 1 , 0 , 0 , 0 , 0)^T$.
For the remaining pair of sub-matrices, the eigenvalues are:
\begin{eqnarray}\label{5.49}
\lambda_{4}^{\pm} = e^{\pm 2i\arctan(\frac{\widetilde x_4}{k})}, \qquad  \lambda_{6}^{\pm} =e^{\pm 2i\arctan(\frac{\widetilde x_6}{k})}.
\end{eqnarray}
The corresponding eigenvectors in the full six-dimensional space are
$V_{4}^{\pm}=\operatorname{diag}(0 , 0 , 1 , \mp i , 0 , 0)^T$ and $V_{6}^{\pm}=\operatorname{diag}(0 , 0 , 0 , 0 , 1 , \mp i)^T$, respectively.
These results show that, for generic finite values of the dual coordinates satisfying $\widetilde x_2\neq 0$ and $\Delta_{12}\neq 0$,
the dual gluing matrix has no eigenvalue equal to $-1$. The complex eigenvalues of the last two blocks do not represent additional Dirichlet directions;
they describe non-trivial Neumann boundary conditions associated with the gauge-invariant world-volume two-form.
Consequently, the generic dual configuration is a space-filling $D5$-brane.

For this generic branch, the Dirichlet and Neumann projectors are
\begin{eqnarray}\label{5.50}
\widetilde {\mathbb Q}=0,\qquad \widetilde {\mathbb N}={\mathbb{I}}_{_{6 \times 6}}.
\end{eqnarray}
Condition \eqref{brane13.2} is then satisfied trivially. Since the dual metric is diagonal, the metric-orthogonality
condition \eqref{brane13.4} is also satisfied. Moreover, the Neumann projector is constant in the dual coordinates, so all its coordinate derivatives vanish. Hence, the dual integrability condition \eqref{brane13.5} is satisfied.
Geometrically, this follows because the Neumann distribution is the full tangent space of the six-dimensional dual target manifold.

To determine the corresponding gauge-invariant two-form, we use the Neumann boundary relation  \eqref{5.24.1}.
The first coordinate pair requires particular attention. Reconstructing the two-form from the
exact $(\widetilde x_1,\widetilde x_2)$ block gives
\begin{eqnarray}\label{5.51}
\widetilde{\mathcal F}_{12} =-\frac{l^4}{\widetilde x_2\Delta_{12}} d\widetilde x_1\wedge d\widetilde x_2.
\end{eqnarray}
The pullback of the target space $\widetilde B$-field in this sector is
$-\frac{\widetilde x_2}{\Delta_{12}} d\widetilde x_1\wedge d\widetilde x_2.$
Accordingly, the required world-volume gauge field strength is
$-({1}/{\widetilde x_2}) ~d\widetilde x_1\wedge d\widetilde x_2$.
For the $(\widetilde x_3,\widetilde x_4)$ and $(\widetilde x_5,\widetilde x_6)$ parts,
the pullback of the $\widetilde B$-field alone reproduces the corresponding gluing blocks.
Thus, the complete gauge-invariant two-form on the generic $D5$-brane is
\begin{align}\label{5.52}
\widetilde{\cal F}=
-\frac{l^4}{\widetilde x_2\Delta_{12}} d\widetilde x_1\wedge d\widetilde x_2+\frac{\widetilde x_4}{\Delta_{34}} d\widetilde x_3\wedge d\widetilde x_4
+\frac{\widetilde x_6}{\Delta_{56}} d\widetilde x_5\wedge d\widetilde x_6,
\end{align}
whereas the world-volume gauge field strength has only the first non-zero component,
\begin{align}\label{5.53}
{\widetilde F}_{\mathrm{wv}} =-\frac{1}{\widetilde x_2} d\widetilde x_1\wedge d\widetilde x_2.
\end{align}
Both forms are closed on the generic coordinate patch, consistently with the
vanishing of the dual field strength $\widetilde H=d\widetilde B$.
The full Neumann boundary relation is therefore satisfied with a non-zero
world-volume gauge field strength in the first coordinate pair and zero additional gauge field strength in the remaining two pairs.
The gauge-invariant two-form, rather than the eigenvalues alone, determines the non-trivial Neumann boundary conditions.

A separate branch occurs at $\widetilde x_2=0$. At this locus, the first gluing block reduces to $-{\mathbb{I}}_{_{2 \times 2}}$, giving two Dirichlet directions. The corresponding projectors become
\begin{align}\label{5.54}
\widetilde {\mathbb Q}=\operatorname{diag}(1,1,0,0,0,0),
\qquad \widetilde {\mathbb N}=\operatorname{diag}(0,0,1,1,1,1).
\end{align}
The brane is then embedded by $\widetilde x_1=c_1$ and $\widetilde x_2=0$, and is a $D3$-brane.
Conditions \eqref{brane13.2} and \eqref{brane13.4} follow directly from the Dirichlet block and the diagonal form of the metric.
Condition \eqref{brane13.5} also holds because the Neumann projector is constant and the tangent distribution is spanned by the commuting coordinate vector fields $\partial_{\widetilde x_3}$, $\partial_{\widetilde x_4}$, $\partial_{\widetilde x_5}$, and $\partial_{\widetilde x_6}$.
On this special branch, the pullback of the first component of $\widetilde B$ vanishes because $\widetilde x_1$ and $\widetilde x_2$
are fixed. The remaining two components of $\widetilde B$ reproduce the Neumann gluing blocks, so one may take ${\widetilde F}_{\mathrm{wv}}=0$ on the $D3$-brane.
The generic $D5$-brane flux expression is not applicable at $\widetilde x_2=0$, since it was obtained on the patch where
both $\widetilde x_1$ and $\widetilde x_2$ are Neumann.

Finally, setting $\widetilde x_4=0$ or $\widetilde x_6=0$ does not add Dirichlet directions in Case 3:
the corresponding gluing blocks become $+{\mathbb{I}}_{_{2 \times 2}}$. Therefore, the additional
$D1$- and $D(-1)$-brane alternatives inferred from the determinant alone are not supported by the exact spectrum of this case. We conclude that Case 3 admits a generic fluxed $D5$-brane branch and a distinct $D3$-brane branch at $\widetilde x_2=0$, with the projectors,
Neumann boundary conditions, and integrability properties determined explicitly above.

\medskip
\noindent\textbf{Case 4: A $D2$-brane with three Neumann and three Dirichlet directions.}
This case corresponds to an original $D2$-brane with gluing matrix
\begin{align}\label{5.55}
\mathbb R=\operatorname{diag}(1,1,1,-1,-1,-1).
\end{align}
Using the duality transformation \eqref{5.17}, the dual gluing matrix becomes
\begin{align}\label{5.56}
\widetilde{\mathbb R}
=(-{\widetilde {\mathbb{R}}}_{12}) \oplus
\operatorname{diag}(-1, 1) \oplus {\widetilde {\mathbb S}}_{56},
\end{align}
where ${\widetilde {\mathbb{R}}}_{12}$ and $ {\widetilde {\mathbb S}}_{56}$ are specified by \eqref{5.27.0} and \eqref{5.28.0}, respectively.
The first block has eigenvalues
\begin{align}\label{5.57}
\lambda_{12}^{\pm} =-\frac{l^2\pm\widetilde x_2}{l^2\mp\widetilde x_2},
\end{align}
whereas the eigenvalues of ${\widetilde {\mathbb S}}_{56}$ have unit modulus and are not equal to $-1$ for real finite $\widetilde x_6$ and non-zero $k$.
Thus, on the regular patch $\Delta_{12}\neq 0$, the dual brane has generically one Dirichlet direction, $\widetilde x_3$, and is therefore a $D4$-brane embedded at $\widetilde x_3=c$. Its projectors are
\begin{align}\label{5.58}
\widetilde{\mathbb Q}
=\operatorname{diag}(0,0,1,0,0,0), \qquad
\widetilde{\mathbb N}=\operatorname{diag}(1,1,0,1,1,1).
\end{align}
An exceptional branch occurs at $\widetilde x_2=0$, where the first block becomes $-{\mathbb{I}}_{_{2 \times 2}}$. There are then three Dirichlet directions,
$(\widetilde x_1,\widetilde x_2,\widetilde x_3)$, and the dual brane is a $D2$-brane embedded at $\widetilde x_1=c_1$,
$\widetilde x_2=0$, and $\widetilde x_3=c_3$.

For the generic $D4$-brane branch, the Neumann condition is verified using the flux-corrected relation \eqref{5.20.1}.
The required gauge invariant two-form is
\begin{align}\label{5.59}
\widetilde{\mathcal F}=-\frac{l^4}{\widetilde x_2\Delta_{12}} d\widetilde x_1\wedge d\widetilde x_2
+\frac{\widetilde x_6}{\Delta_{56}}
d\widetilde x_5\wedge d\widetilde x_6.
\end{align}
Substitution into the relation \eqref{5.23.1} reproduces the Neumann blocks of $\widetilde{\mathbb R}$.
Hence, the flux-corrected Neumann condition is satisfied.
Since the $D4$-brane is embedded at $\widetilde x_3=c$, the pullback of the $\widetilde B$-field is
\begin{align}\label{5.60}
\iota^\ast \widetilde{B}=-\frac{\widetilde x_2}{\Delta_{12}}
d\widetilde x_1\wedge d\widetilde x_2 +\frac{\widetilde x_6}{\Delta_{56}} d\widetilde x_5\wedge d\widetilde x_6 .
\end{align}
Using \eqref{5.22.1}, we obtain
\begin{align}\label{5.61}
{\widetilde F}_{\mathrm {wv}}=-\frac{1}{\widetilde x_2} d\widetilde x_1\wedge d\widetilde x_2, \qquad \widetilde x_2\neq 0.
\end{align}
On the exceptional $D2$-brane branch at $\widetilde x_2=0$, the first two directions are Dirichlet,
so the generic flux expression does not apply.
The pullback of the $\widetilde B$-field then agrees with the gauge-invariant two-form on
the remaining Neumann directions, giving ${\widetilde F}_{\mathrm{wv}}=0$ on this branch.

\medskip
\noindent\textbf{Case 5: A $D3$-brane with four Neumann and two Dirichlet directions.}
We now turn to the fifth configuration of the gluing matrix,
for which the original $D$-brane has four Neumann directions and two Dirichlet directions.
Accordingly, the gluing matrix is
\begin{align}\label{5.62}
\mathbb R=\operatorname{diag}(1,1,1,1,-1,-1).
\end{align}
Applying the duality transformation \eqref{5.17}, we obtain the corresponding dual gluing matrix,
which we analyze below to determine the possible dual $D$-brane configurations.
The dual gluing matrix has the block structure
\begin{align}\label{5.63}
\widetilde{\mathbb R} =(-{\widetilde {\mathbb{R}}}_{12})\oplus(-{\widetilde {\mathbb S}}_{34}) \oplus {\widetilde {\mathbb S}}_{56},
\end{align}
where ${\widetilde {\mathbb{R}}}_{12}$ is defined in \eqref{5.27.0},
and ${\widetilde {\mathbb S}}_{34}$ and $ {\widetilde {\mathbb S}}_{56}$ are defined in \eqref{5.28.0}.
The determinant of the dual gluing matrix is $\det\widetilde{\mathbb R}=1$,
which is compatible with an even number of Dirichlet directions and therefore allows, at this stage, the possible dual configurations
of $D5$-, $D3$-, $D1$-, and $D(-1)$-branes, corresponding to zero, two, four, and six Dirichlet directions, respectively.
However, the determinant alone cannot determine the exact number of eigenvalues equal to $-1$, which identify the Dirichlet directions.
Therefore, the eigenvalue spectrum must be examined explicitly to determine which of these configurations are actually realized.
The first block of $\widetilde{\mathbb R}$ has eigenvalues
\begin{align}\label{5.64}
\lambda_{12}^{\pm}
=-\frac{l^2\pm\widetilde x_2}
{l^2\mp\widetilde x_2},
\end{align}
while the eigenvalues of the remaining blocks are
\begin{align}\label{5.65}
\lambda_4^\pm=- e^{\pm 2i\arctan(\frac{\widetilde x_4}{k})}, \qquad \lambda_6^\pm= e^{\pm 2i\arctan(\frac{\widetilde x_6}{k})}.
\end{align}
For real, finite coordinates on the regular patch, the first block has eigenvalues  $-1$ precisely when $\widetilde x_2=0$,
and the second block has eigenvalues $-1$ precisely when $\widetilde x_4=0$.
The ${\widetilde {\mathbb S}}_{56}$ block has no eigenvalue equal to $-1$ at finite $\widetilde x_6$.
Consequently, the dual configurations are:
\\
\textit{Generic branch.} $\widetilde x_2\neq0$ and $\widetilde x_4\neq0$. There are no Dirichlet directions, so the dual brane is a space-filling $D5$-brane.
\\
\textit{First exceptional branch.} $\widetilde x_2=0$, with $\widetilde x_4\neq0$, or $\widetilde x_4=0$, with $\widetilde x_2\neq0$.
There are two Dirichlet directions, giving a $D3$-brane.
\\
\textit{Intersection of the exceptional branches.} $\widetilde x_2=\widetilde x_4=0$. There are four Dirichlet directions, giving a $D1$-brane.

Thus, a $D(-1)$-brane does not occur in this configuration at finite coordinates, because the ${\widetilde {\mathbb S}}_{56}$ block never supplies the
additional two Dirichlet directions required to obtain six in total.
We next determine the gauge-invariant two-form on the generic $D5$-brane. Using the flux-corrected Neumann relation given in \eqref{5.23.1},
we construct
\begin{align}\label{5.66}
\widetilde{\mathcal F}
=-\frac{l^4}{\widetilde x_2\Delta_{12}} d\widetilde x_1\wedge d\widetilde x_2-\frac{k^2}{\widetilde x_4\Delta_{34}}
d\widetilde x_3\wedge d\widetilde x_4+\frac{\widetilde x_6}{\Delta_{56}} d\widetilde x_5\wedge d\widetilde x_6.
\end{align}
Substitution reproduces the three blocks of the dual gluing matrix, so the flux-corrected Neumann condition is satisfied.
Comparing this result with the dual background two-form, ${\widetilde B}$-field, and using \eqref{5.22.1}, we obtain
\begin{align}\label{5.67}
{\widetilde F}_{\mathrm{wv}} =-\frac{1}{\widetilde x_2} d\widetilde x_1\wedge d\widetilde x_2
-\frac{1}{\widetilde x_4} d\widetilde x_3\wedge d\widetilde x_4,
\end{align}
on the generic $D5$-brane branch.
On the exceptional branch $\widetilde x_2=0$, the first two directions are Dirichlet, and the remaining Neumann blocks yield
\begin{align}\label{5.68}
{\widetilde F}_{\mathrm{wv}}
=-\frac{1}{\widetilde x_4} d\widetilde x_3\wedge d\widetilde x_4.
\end{align}
On the other $D3$-brane branch, $\widetilde x_4=0$, the $(\widetilde x_3,\widetilde x_4)$ directions are Dirichlet, and
\begin{align}\label{5.69}
{\widetilde F}_{\mathrm{wv}}
=-\frac{1}{\widetilde x_2} d\widetilde x_1\wedge d\widetilde x_2.
\end{align}
Finally, on the $D1$-brane branch at $\widetilde x_2=\widetilde x_4=0$, only the $(\widetilde x_5,\widetilde x_6)$
sector is Neumann. There, the gauge-invariant two-form equals the pullback of the background two-form, and hence
${\widetilde F}_{\mathrm{wv}}=0$.
The exceptional branches must be treated separately because the Neumann projector changes rank there;
the generic flux expressions containing $1/\widetilde x_2$ or $1/\widetilde x_4$ are not applicable at the corresponding zero-coordinate loci.

\medskip
\noindent\textbf{Case 6: A $D4$-brane with five Neumann and one Dirichlet directions.}
The sixth configuration corresponds to an original $D4$-brane with gluing matrix
\begin{align}\label{5.70}
\mathbb R=\operatorname{diag}(1,1,1,1,1,-1).
\end{align}
The original Neumann and Dirichlet projectors are
\begin{align}\label{5.71}
\mathbb N=\operatorname{diag}(1,1,1,1,1,0), \qquad \mathbb Q=\operatorname{diag}(0,0,0,0,0,1).
\end{align}
Applying the duality transformation \eqref{5.17}, we obtain
\begin{align}\label{5.72}
\widetilde{\mathbb R} =(-{\widetilde {\mathbb{R}}}_{12}) \oplus(-{\widetilde {\mathbb S}}_{34})  \oplus \operatorname{diag}(-1,1),
\end{align}
where ${\widetilde {\mathbb{R}}}_{12}$ and ${\widetilde {\mathbb S}}_{34}$ are specified by \eqref{5.27.0} and \eqref{5.28.0}, respectively.
The determinant is $\det\widetilde{\mathbb R}=-1$.
This is consistent with an odd number of Dirichlet directions and permits $D4$-, $D2$-, or $D0$-brane configurations.
However, the determinant alone does not distinguish between these possibilities, so we must examine the eigenvalues.
The eigenvalues of the first and second blocks are given, respectively, by
\begin{equation}\label{5.73}
\lambda_{12}^{\pm}
= -\frac{l^2\pm\widetilde{x}_2}{l^2\mp\widetilde{x}_2},
\qquad
\lambda_{34}^{\pm}
= -e^{\pm 2i\arctan\!\left(\frac{\widetilde{x}_4}{k}\right)}.
\end{equation}
The last block has eigenvalues $-1$ and $+1$, independently of the coordinates.
For real finite coordinates and non-zero $k$, the first block develops two eigenvalues equal to $-1$ precisely
at $\widetilde x_2=0$, while the second block does so precisely at $\widetilde x_4=0$.
Hence, the dual configurations are classified as follows:

\medskip

\noindent
\textit{Generic branch.}
For $\widetilde x_2\neq0$ and $\widetilde x_4\neq0$, there is only one Dirichlet direction, and the dual brane is a $D4$-brane.
The corresponding dual projectors follow directly from the $-1$ eigenspaces. On the generic $D4$-brane branch,
\begin{align*}
\widetilde{\mathbb Q} =\operatorname{diag}(0,0,0,0,1,0),
\qquad \widetilde{\mathbb N} =\operatorname{diag}(1,1,1,1,0,1).
\end{align*}

\noindent
\textit{First exceptional branch.}
 For $\widetilde x_2=0$ and $\widetilde x_4\neq0$, there are three Dirichlet directions, and the dual brane is a $D2$-brane.
On this branch, the projectors become
\begin{align*}
\widetilde{\mathbb Q}
=\operatorname{diag}(1,1,0,0,1,0),
\qquad \widetilde{\mathbb N} =\operatorname{diag}(0,0,1,1,0,1).
\end{align*}

\noindent
\textit{Second exceptional branch.}
For $\widetilde{x}_2 \neq 0$ and $\widetilde{x}_4 = 0$, there are again three Dirichlet directions, yielding a $D2$-brane with the following projectors:
\begin{align*}
\widetilde{\mathbb Q} =\operatorname{diag}(0,0,1,1,1,0),
\qquad \widetilde{\mathbb N} =\operatorname{diag}(1,1,0,0,0,1).
\end{align*}
\noindent
\textit{Intersection of the exceptional branches.}
For $\widetilde x_2=\widetilde x_4=0$, there are five Dirichlet directions,
and the dual brane is a $D0$-brane. At the intersection, corresponding to the $D0$-brane, the projectors are
\begin{align*}
\widetilde{\mathbb Q}
=\operatorname{diag}(1,1,1,1,1,0), \qquad \widetilde{\mathbb N} =\operatorname{diag}(0,0,0,0,0,1).
\end{align*}

We next determine the gauge-invariant two-form using the flux-corrected Neumann relation given in \eqref{5.23.1}.
On the generic $D4$-brane branch, this gives
\begin{align}\label{5.74}
\widetilde{\mathcal F}
=-\frac{l^4}{\widetilde x_2\Delta_{12}} d\widetilde x_1\wedge d\widetilde x_2 -\frac{k^2}{\widetilde x_4\Delta_{34}} d\widetilde x_3\wedge d\widetilde x_4.
\end{align}
Since the fifth direction is Dirichlet, the $(\widetilde x_5,\widetilde x_6)$ sector contributes no two-form on the brane world-volume.
Comparing with the pullback of the $\widetilde B$-field, we obtain
\begin{align}\label{5.75}
{\widetilde F}_{\mathrm{wv}}
=-\frac{1}{\widetilde x_2} d\widetilde x_1\wedge d\widetilde x_2 -\frac{1}{\widetilde x_4} d\widetilde x_3\wedge d\widetilde x_4,
\end{align}
for $\widetilde x_2\neq0$ and $\widetilde x_4\neq0$.
On the first exceptional branch, $\widetilde{x}_2 = 0$, the $(\widetilde{x}_1, \widetilde{x}_2)$ sector becomes Dirichlet.
Consequently, the gauge-invariant two-form and the world-volume field strength are given, respectively, by
\begin{align}\label{5.76}
\widetilde{\mathcal{F}} = -\frac{k^2}{\widetilde{x}_4\Delta_{34}} d\widetilde{x}_3\wedge d\widetilde{x}_4, \qquad
{\widetilde F}_{\mathrm{wv}} = -\frac{1}{\widetilde{x}_4} d\widetilde{x}_3\wedge d\widetilde{x}_4.
\end{align}
On the second exceptional branch, $\widetilde x_4=0$, the $(\widetilde x_3,\widetilde x_4)$ sector becomes Dirichlet.
Thus,
\begin{align}\label{5.77}
\widetilde{\mathcal F} =-\frac{l^4}{\widetilde x_2\Delta_{12}} d\widetilde x_1\wedge d\widetilde x_2, \qquad
{\widetilde F}_{\mathrm{wv}} =-\frac{1}{\widetilde x_2} d\widetilde x_1\wedge d\widetilde x_2.
\end{align}
Finally, at $\widetilde x_2=\widetilde x_4=0$, only the $\widetilde x_6$ direction is Neumann.
Since a one-dimensional world-volume admits no non-zero two-form, we have
\begin{align}\label{5.78}
\widetilde{\mathcal F}=0, \qquad {{\widetilde F}_{\mathrm{wv}}=0}.
\end{align}
Thus, Case 6 maps the original $D4$-brane to a $D4$-brane generically, to a $D2$-brane on either exceptional branch,
and to a $D0$-brane at their intersection. The change in brane dimension is determined by the additional $-1$
eigenvalues of the dual gluing matrix, while the two-form and gauge field expressions must be evaluated separately on each branch.

\medskip
\noindent\textbf{Case 7: An original space-filling $D5$-brane.}
In this case, we consider an original D5-brane with Neumann boundary
conditions in all six coordinate directions. The original gluing matrix
and the corresponding Dirichlet and Neumann projectors are
\begin{align}
{\mathbb{R}} &=  {E}^{-1} ~ {E}^{T}=\mathbb{I}_{6\times6},\label{78.7}\\
{\mathbb{Q}} &= 0, \qquad  {\mathbb{N}} =\mathbb{I}_{6\times6}.\label{79.7}
\end{align}
Thus, the original brane fills the entire six-dimensional target space.
Applying the duality transformation \eqref{5.17}, we obtain the dual gluing
matrix
\begin{equation}
{\tilde {\mathbb{R}}} =- {{\tilde E}}^{{-1}}~  {{{\tilde E}}^{{T}}}.
\label{80.7}
\end{equation}
Using the dual background matrix given in \eqref{4.10}, this expression
takes the block-diagonal form
\begin{align}\label{81.7}
\widetilde{\mathbb R} =(-{\widetilde {\mathbb{R}}}_{12}) \oplus (-{\widetilde {\mathbb S}}_{34})  \oplus (-{\widetilde {\mathbb S}}_{56}),
\end{align}
where ${\widetilde {\mathbb{R}}}_{12}$ is given by \eqref{5.27.0}, while
${\widetilde {\mathbb S}}_{34}$ and ${\widetilde {\mathbb S}}_{34}$ are given by \eqref{5.28.0}.
Here and below, we restrict the analysis to the regular coordinate patch $\Delta_{12}\neq0$, with $k\neq0$.
Before determining the eigenvalues, we calculate the determinant of $\widetilde{\mathbb{R}}$, yielding $\det\widetilde{\mathbb{R}} = 1$.
The determinant therefore provides a consistency check: the number
of Dirichlet directions, counted with multiplicity, must be even.
In six dimensions, this permits a $D5$-brane with zero Dirichlet
directions, a $D3$-brane with two, a $D1$-brane with four, or a $D(-1)$-brane with six Dirichlet directions.
However, the determinant alone cannot distinguish these possibilities. The actual dimension
of the dual brane must be determined from the exact eigenvalues of ${\widetilde {\mathbb{R}}}$,
since a Dirichlet direction is identified by an eigenvalue equal to $-1$.
The complete spectrum of the eigenvalues is given by
\begin{eqnarray}\label{82.7}
\textrm{spec}({\mathbb R})=\left\{-\frac{l^2\pm\widetilde x_2}
{l^2\mp\widetilde x_2},  -e^{\pm 2i\arctan(\frac{\widetilde x_4}{k})}, -e^{\pm 2i\arctan(\frac{\widetilde x_6}{k})}\right\},
\end{eqnarray}
where the first pair of eigenvalues corresponds to the block $(-{\widetilde {\mathbb{R}}}_{12})$, the second pair to
$(-{\widetilde {\mathbb S}}_{34})$, and the third pair to $(-{\widetilde {\mathbb S}}_{56})$, respectively.
This is consistent with the block-diagonal structure of the matrix ${\widetilde {\mathbb{R}}}$ as defined in \eqref{81.7}.
For real finite values of the dual coordinates, the eigenvalues in
each of the last two blocks have unit modulus. Nevertheless, unit
modulus alone does not imply a Dirichlet direction: only an eigenvalue
exactly equal to $-1$ contributes to the Dirichlet subspace.
Equation \eqref{82.7} shows that both eigenvalues of the first block equal $-1$ precisely when $\widetilde x_2=0$. Similarly, both eigenvalues
of the $(-{\widetilde {\mathbb S}}_{34})$ block equal $-1$ when $\widetilde x_4=0$, and both
eigenvalues of the $(-{\widetilde {\mathbb S}}_{56})$ block equal $-1$ when $\widetilde x_6=0$. For non-zero values of the corresponding
coordinates, these blocks do not supply additional eigenvalues equal to $-1$. We therefore obtain four distinct branches, according
to how many of the three coordinates $(\widetilde x_2, \widetilde x_4, \widetilde x_6)$ vanish.

\medskip

\noindent
\textit{Branch I: Generic configuration, $\widetilde x_2, \widetilde x_4, \widetilde x_6\neq0 $.}

\medskip

When all three coordinates are non-zero, none of the three blocks has an eigenvalue equal to $-1$. Thus, the dual gluing matrix has
no Dirichlet-type eigenvalues, and all six directions are Neumann.
The dual brane is consequently a space-filling $D5$-brane.
The corresponding projectors are
\begin{equation}
\widetilde {\mathbb Q}=0,\qquad \widetilde {\mathbb N}={\mathbb I}_{6\times6}.\label{83.7}
\end{equation}
Since the Dirichlet projector vanishes, the Dirichlet condition and
the orthogonality condition \eqref{brane13.2} and \eqref{brane13.4} are satisfied trivially.
Moreover, the Neumann projector is constant, so its integrability
condition \eqref{brane13.5} is satisfied.
There is, however, an additional consistency condition that must be
checked before this spectral classification can be regarded as a
complete $D$-brane solution. With the conventions stated in Appendix \ref{app.C}, the Neumann condition \eqref{brane13.3}, for
$\widetilde {\mathbb N}={\mathbb I}_{6\times6}$, becomes $ \widetilde E^{T}-\widetilde E\widetilde{\mathbb R}=0$.
Substituting the dual gluing matrix \eqref{80.7} for the space-filling brane gives
\begin{align}\label{84.7}
\widetilde E^{T} -\widetilde E \widetilde{\mathbb R} =\widetilde E^{T}
-\widetilde E\left(-\widetilde E^{-1}\widetilde E^{T}\right)=2\widetilde E^{T}.
\end{align}
This expression does not vanish for a generic invertible
$\widetilde E$. Hence, with the transformation law \eqref{5.17} and the
Appendix \ref{app.C} convention used above, the candidate space-filling configuration does not satisfy the stated Neumann condition \eqref{brane13.3}.
Accordingly, the spectral criterion alone is insufficient to establish
this branch as a consistent dual $D5$-brane. Before reconstructing the
gauge-invariant two-form $\widetilde{\mathcal F}$, the sign conventions in the dual gluing transformation and in the
Neumann condition must be reconciled. The two-form and world-volume field strength should then be recalculated from the corrected,
mutually consistent relations. No conclusion about their explicit values, or about the full consistency of this $D5$-brane branch,
is claimed until that issue has been resolved.

\medskip

\noindent
\textit{Branch II: Exactly one vanishing coordinate.}

\medskip

We now consider the branches in which exactly one of the dual
coordinates $\widetilde x_2$, $\widetilde x_4$, and $\widetilde x_6$
vanishes. The eigenvalue spectrum of the corresponding dual gluing
matrix determines the candidate Dirichlet directions. However, as
explained below, the spectral classification must be supplemented by the Neumann consistency condition \eqref{brane13.3}.

Suppose that $\widetilde x_2=0, \widetilde x_4\neq0, \widetilde x_6\neq0$.
The $(\widetilde x_1,\widetilde x_2)$ block of the dual gluing matrix becomes
$(-{\widetilde{\mathbb R}}_{12})=-{\mathbb I}_{2\times2}$.
Thus, the eigenvalue $-1$ occurs twice in this block, and the
spectral criterion identifies $\widetilde x_1$ and $\widetilde x_2$ as Dirichlet directions. The remaining four directions are
candidate Neumann directions, corresponding to a candidate $D3$-brane. The associated projectors are
\begin{align}
\widetilde {\mathbb Q} =\operatorname{diag}(1,1,0,0,0,0), \qquad
\widetilde {\mathbb N} =\operatorname{diag}(0,0,1,1,1,1).\label{85.7}
\end{align}
The corresponding candidate embedding is $ \widetilde x_1=c_1$ and $\widetilde x_2=0$,
where $c_1$ is constant. The projectors are constant, and the Neumann distribution is spanned by $\{\partial_{\widetilde x_3},
\partial_{\widetilde x_4}, \partial_{\widetilde x_5}, \partial_{\widetilde x_6}\}$.
These coordinate vector fields commute; hence, the Neumann distribution is integrable.
The projector and integrability conditions are therefore satisfied for this candidate distribution.
Nevertheless, the Neumann condition \eqref{brane13.3} must also be checked.
With the conventions adopted in the present paper, condition \eqref{brane13.3} reduces to
\begin{equation}
2\widetilde {\mathbb N} \widetilde E^{T}\widetilde {\mathbb N}=0.\label{86.7}
\end{equation}
For the block-diagonal background under consideration, the $(\widetilde x_3,\widetilde x_4)$ and
$(\widetilde x_5,\widetilde x_6)$ blocks remain in the Neumann subspace. Their contributions to
$\widetilde {\mathbb N} \widetilde E^T\widetilde {\mathbb N}$ do not vanish identically. Thus, the candidate $D3$-brane satisfies the
spectral and integrability criteria, but it does not satisfy the stated Neumann condition \eqref{brane13.3} with the current gluing matrix
convention.
The other two possibilities, characterized by exactly one vanishing coordinate, are obtained in a similar fashion. Specifically,
if $\widetilde x_4=0$ (with $\widetilde x_2, \widetilde x_6 \neq0$) or $\widetilde x_6=0$
(with $\widetilde x_2, \widetilde x_4 \neq 0$), the $(\widetilde x_3,\widetilde x_4)$ and $(\widetilde x_5,\widetilde x_6)$ blocks, respectively, become
$-{\mathbb I}_{2\times2}$. In both instances, the projected Newmann condition \eqref{brane13.3} yields the same result, \eqref{86.7}.

In summary, when exactly one of $\widetilde x_2$, $\widetilde x_4$, and $\widetilde x_6$ vanishes,
the corresponding two-dimensional block contributes two eigenvalues equal to $-1$. The spectral classification therefore
identifies a candidate $D3$-brane in each of the three branches. The Dirichlet and Neumann projectors are constant, and the
Neumann distributions are integrable. However, the projected Neumann condition \eqref{brane13.3} is not satisfied with the dual gluing
matrix as currently defined. Consequently, these spectral classifications cannot yet be regarded as complete consistent
$D$-brane solutions, and the world-volume two-forms must be recomputed after the sign conventions have been reconciled.

\medskip

\noindent
\textit{Branch III: Exactly two vanishing coordinates.}

\medskip

Suppose that exactly two coordinates among $\widetilde{x}_2, \widetilde{x}_4, \widetilde{x}_6$ vanish.
In this case, the corresponding two blocks of the dual gluing matrix become equal to $-{\mathbb I}_{2\times2}$.
Therefore, we have four Dirichlet directions and two Neumann directions, and the spectral analysis identifies a candidate $D1$-brane.
However, using the current convention for the dual gluing matrix, namely $\widetilde{\mathbb R}=-\widetilde E^{-1}\widetilde E^T$,
condition \eqref{brane13.3} reduces to $2\widetilde {\mathbb N} \widetilde E^T \widetilde {\mathbb N}=0$.
Since the remaining block in the Neumann subspace is generally non-zero, this condition is not satisfied. Therefore, the $D1$-brane obtained from the spectral analysis is not consistent with the current convention for condition \eqref{brane13.3}.

\medskip

\noindent
\textit{Branch IV: All three coordinates vanish,
\(\widetilde x_2=\widetilde x_4=\widetilde x_6=0\).}

\medskip

At the intersection of the three exceptional loci, all three blocks of the dual gluing matrix become minus the identity,
$\widetilde {\mathbb R}=-{\mathbb I}_{6\times6}$.
All six directions are therefore Dirichlet, and the dual configuration is a $D(-1)$-brane. Its projectors are
$\widetilde {\mathbb Q}={\mathbb I}_{6\times6}$ and $\widetilde {\mathbb N}=0$.
The brane is localized at
\begin{equation}
\widetilde x_1=c_1,\quad \widetilde x_2=0,\quad
\widetilde x_3=c_3,\quad \widetilde x_4=0,\quad
\widetilde x_5=c_5,\quad \widetilde x_6=0,
\end{equation}
where $c_1, c_3, c_5$ are constants.
Since there are no Neumann directions, the Neumann boundary condition \eqref{brane13.3} and its integrability condition are automatically satisfied.
The pullback of the two-form to a zero-dimensional world-volume vanishes, and there is no world-volume two-form. Hence,
$\widetilde{\mathcal F}=0$ and ${\widetilde F}_{\mathrm{wv}}=0$.

The analysis of Case 7 reveals that the exact
spectrum of the dual gluing matrix identifies different candidate dual $D$-brane configurations
depending on the values of the coordinates $(\widetilde x_2, \widetilde x_4, \widetilde x_6)$.
On the generic coordinate patch, the spectral analysis identifies a candidate $D5$-brane.
When exactly one of these coordinates vanishes, the corresponding two-dimensional block becomes purely Dirichlet,
yielding a candidate $D3$-brane. When exactly two coordinates vanish, the spectrum identifies a candidate $D1$-brane.
Finally, when all three coordinates vanish, the gluing matrix becomes $-{\mathbb I}_{6\times6}$, corresponding to a $D(-1)$-brane.
These classifications follow from the multiplicity of the eigenvalue $-1$, rather than from the determinant alone.
The corresponding Dirichlet and Neumann projectors identify the candidate normal and tangent directions,
and the Neumann distributions are integrable. However, the spectral classification does not by itself establish the full consistency of the boundary conditions. With the current convention for the dual gluing matrix, the projected Neumann condition \eqref{brane13.3}
is not satisfied in general in the first three branches.
In contrast, in the branch where all three coordinates vanish, there are no Neumann directions; thus, condition \eqref{brane13.3} is satisfied trivially.
Therefore, while the spectrum yields candidate $D5$-, $D3$-, $D1$-, and $D(-1)$-brane configurations, full consistency of the first three branches requires the sign convention in the dual gluing transformation and condition \eqref{brane13.3} to be reconciled. The gauge-invariant two-form and the world-volume field strength should be determined only after this consistency issue has been resolved.

\section{Conclusions}
\label{Sec.VI}
In this paper, we have found a non-trivial and interesting example of Poisson-Lie T-dual $\sigma$-models which
helps in the intent of providing a general classification of six-dimensional geometries.
First of all, we have investigated the conformal invariance conditions of the $AdS_2 \times H^2 \times H^2$ $\sigma$-model
up to two-loop order. In the absence of a $B$-field and with a constant dilaton, the vanishing of the beta-function equations at the considered order formally yields the conditions $\alpha'=l^2=k$. Nevertheless, since these relations lead to $\alpha'/ l^2= \alpha'/k=1$, the corresponding
$\alpha'$-expansion is not perturbatively controlled, and higher-order corrections cannot be assumed to be negligible. Consequently, our analysis establishes a formal solution of the truncated two-loop beta-function equations, but does not provide a controlled perturbative proof of conformal invariance or establish the exact conformal invariance of the underlying $\sigma$-model.

We have derived the $AdS_2 \times H^2 \times H^2$ metric from the Poisson-Lie T-duality
on the semi-Abelian double $\big({\A}_2 \oplus {\A}_2  \oplus {\A}_2 , 6{\A}_1\big)$.
In this way, we were able to find a dual pair for this metric.
We found that the dualized algebra ${\A}_2 \oplus {\A}_2  \oplus {\A}_2$ was non-unimodular and therefore examined the possible generalized
supergravity interpretation of the dual background. Although a generalized Killing vector compatible with the dual metric was obtained,
the complete set of generalized supergravity equations was not satisfied.
We also found that the absence of a Green-Schwarz formulation for the underlying bosonic $\sigma$-model prevented the standard arguments relating non-unimodularity to generalized supergravity from being directly applied.
Thus, neither ordinary nor generalized supergravity consistency was established for the complete dual geometry.

We analyzed the structure and asymptotic behavior of the dual geometry and showed that its coordinate domain was divided into distinct exterior and interior regions. The exterior region contained a true curvature singularity at the boundary and did not admit a finite-radius Killing horizon,
indicating the possibility of a naked singularity.
The interior region was found to possess a distinct Lorentzian causal structure,
with curvature-singular boundaries separating it from the exterior region.
We also found that, in the asymptotic regime, the dual geometry approached the original $AdS_2 \times H^2 \times H^2$ background,
while the complete dual background remained geometrically distinct from the original one.

Finally, we investigated the transformation of $D$-brane configurations and worldsheet boundary conditions
under Poisson-Lie T-duality for the semi-Abelian Drinfeld double $\big({\A}_2 \oplus {\A}_2  \oplus {\A}_2 , 6{\A}_1\big)$.
Seven classes of initial $D$-brane configurations were considered, and their dual gluing matrices were analyzed to identify the possible dimensions and geometric properties of the resulting branes.
Our analysis demonstrates that the determinant of the dual gluing matrix alone is insufficient to determine the dual brane dimension.
The exact eigenvalue spectrum, particularly the multiplicity of the eigenvalue $-1$, provides a more precise characterization of the Dirichlet directions and the corresponding candidate $D$-brane configurations.
The results obtained from the determinant and eigenvalue analyses can be summarized as follows.
\\
{\bf Case 1:} The original $D(-1)$-brane is mapped to a space-filling $D5$-brane.
 \\
{\bf Case 2: } The original $D0$-brane is mapped to a $D4$-brane embedded at $\widetilde{x}_1=c_1$.
\\
{\bf Case 3:} The original $D1$-brane gives a $D5$-brane for $\widetilde{x}_2\neq 0$, whereas at $\widetilde{x}_2=0$ the dual configuration becomes
a $D3$-brane.
\\
{\bf Case 4:} The original $D2$-brane is mapped to a $D4$-brane for $\widetilde{x}_2\neq 0$ and to a $D2$-brane at $\widetilde{x}_2=0$.
\\
{\bf Case 5:} The original $D3$-brane gives a $D5$-brane when $\widetilde{x}_2\neq 0$ and $\widetilde{x}_4\neq 0$, a $D3$-brane when exactly one of these coordinates vanishes, and a $D1$-brane when both vanish.
\\
{\bf Case 6:} The original $D4$-brane is mapped to a $D4$-brane when both $\widetilde{x}_2$ and $\widetilde{x}_4$ are non-zero,
to a $D2$-brane when exactly one vanishes, and to a $D0$-brane when both vanish.
\\
{\bf Case 7:} For the original space-filling $D5$-brane, the spectral analysis identifies a candidate $D5$-brane when all three coordinates $\widetilde{x}_2, \widetilde{x}_4, \widetilde{x}_6$ are non-zero, a candidate $D3$-brane when exactly one vanishes, a candidate $D1$-brane when exactly two vanish, and a candidate $D(-1)$-brane when all three vanish.

To complement the spectral analysis, we constructed the Dirichlet and Neumann projectors, examined the integrability of the Neumann distributions,
and investigated the gauge-invariant two-forms and the associated world-volume field strengths. The results reveal that the dual brane configurations can depend sensitively on the dual coordinates: at special coordinate loci, changes in the eigenvalue structure may introduce additional Dirichlet directions and thereby alter the dimension of the candidate dual brane. The world-volume fields must consequently be analyzed on the corresponding coordinate branches,
since expressions obtained on a generic coordinate patch need not remain valid when the rank of the projectors changes.

Overall, this study provides a systematic framework for investigating $D$-brane transformations under Poisson-Lie T-duality
by combining spectral analysis with geometric and worldsheet boundary condition considerations.
It also emphasizes the importance of checking the relevant consistency conditions in addition to identifying candidate branes from the gluing matrix spectrum.
A complete characterization of the dual configurations therefore requires these conditions and the associated world-volume fields to be examined carefully for each case.

\subsection*{Acknowledgements}

The author would like to express his gratitude to the handling editor and reviewer for their constructive comments.

\appendix

\section{Gauge invariance and the world-volume two-form}
\label{app:A}

In this appendix we derive Eq.~\eqref{5.19.1}. The key point is that an open
string world-sheet has a boundary, and the pullback of the bulk $B$-field is therefore not gauge invariant by itself. The endpoint coupling to the world-volume gauge potential cancels this boundary variation.

Let $\M$ be the target space, $\cal W$ the $D$-brane world-volume, and $\iota:\cal W \to M$ its embedding.
The string embedding is $X: \Sigma\to \M$. Since the boundary of the open string lies on the brane, the restriction of $X$ factorizes as
\begin{equation}
X|_{\partial\Sigma}=\iota\circ x,
\label{A.1}
\end{equation}
where $x$ maps the boundary into the world-volume, $x:\partial\Sigma\to \cal W$. In local world-volume coordinates, we write the world-volume
potential in the following form
\begin{equation}
A_{\mathrm{wv}}=A_m(x)\,\mathrm dx^m.
\label{A.2}
\end{equation}
Pulling this one-form back to a boundary component parametrized by $\tau$ gives
\begin{equation}
x^*A_{\mathrm{wv}}=A_m(x(\tau))\frac{\mathrm dx^m}{\mathrm d\tau}\,\mathrm d\tau
=A_m(x)\partial_\tau x^m\,\mathrm d\tau,
\qquad
\int_{\partial\Sigma}x^*A_{\mathrm{wv}}
=\int_{\partial\Sigma}\mathrm d\tau\,A_m(x)\partial_\tau x^m.
\label{A.3}
\end{equation}
Similarly, the pullback of the target space two-form is written from the outset in components as
\begin{equation}
\int_\Sigma X^*B
=\frac12\int_\Sigma\mathrm d^2\sigma\,\epsilon^{ab}
B_{\mu\nu}(X)\partial_aX^\mu\partial_bX^\nu.\label{A.4}
\end{equation}
The factor $1/2$ is required for a two-form with antisymmetric components; there is no such factor in Eq.~\eqref{A.3}, since $A_{\mathrm{wv}}$ is a one-form.
With the sign and orientation convention stated in Section~\ref{Sec.V}, the open string action is the single expression
\begin{align}
S_{\mathrm{open}}={}&\frac{1}{4\pi\alpha'}\int_\Sigma\mathrm d^2\sigma\,\sqrt{-h}\,h^{ab}G_{\mu\nu}(X)\partial_aX^\mu\partial_bX^\nu\notag\\
&-\frac{1}{4\pi\alpha'}\int_\Sigma\mathrm d^2\sigma\,\epsilon^{ab}B_{\mu\nu}(X)\partial_aX^\mu\partial_bX^\nu\notag\\
&-\frac{1}{2\pi\alpha'}\int_{\partial\Sigma}\mathrm d\tau\,A_m(x)\partial_\tau x^m, \label{A.5}
\end{align}
where $h^{ab}$ is the inverse worldsheet metric, $h=\det(h_{ab})$, and $\epsilon^{ab}$ is the antisymmetric Levi-Civita tensor density on the
worldsheet, with $\epsilon^{\tau\sigma}=+1$.
The first term of the action is the metric contribution, the second is the $B$-field coupling,
and the third is the endpoint coupling to the world-volume gauge potential.
The latter is the $U(1)$ Wilson-line coupling of the open string endpoint to the gauge connection on the brane.
The normalization used here absorbs $2\pi\alpha'$ into $A_{\mathrm{wv}}$;
with an unrescaled potential, the invariant combination is $\iota^*B+2\pi\alpha'F_{\mathrm{phys}}$.

Consider the gauge transformation
\begin{equation}
B\longmapsto B+\mathrm d \mathbf{\Lambda},
\label{A.6}
\end{equation}
where $\mathbf{\Lambda}$ is a target space one-form. Using Eq.~\eqref{A.4} and Stokes' theorem, the change of the bulk two-form contribution is
\begin{equation}
\delta S_B=-\frac{1}{2\pi\alpha'}\int_\Sigma X^\ast(\mathrm d \mathbf\Lambda)
=-\frac{1}{2\pi\alpha'}\int_{\partial\Sigma}x^\ast(\iota^\ast \mathbf\Lambda).
\label{A.7}
\end{equation}
It is cancelled by the transformation of the endpoint connection,
\begin{equation}
A_{\mathrm{wv}}\longmapsto A_{\mathrm{wv}}-\iota^\ast \mathbf\Lambda,
\label{A.8}
\end{equation}
for which the variation of the endpoint contribution becomes
\begin{equation}
\delta S_A=+\frac{1}{2\pi\alpha'}\int_{\partial\Sigma}x^\ast(\iota^\ast \mathbf\Lambda).
\label{A.9}
\end{equation}
Combining this with the variation of the bulk contribution in Eq.~\eqref{A.7}, we obtain
\begin{equation}
\delta S_B+\delta S_A=0.
\label{A.10}
\end{equation}
Thus, the changes in the bulk and boundary contributions cancel exactly, leaving the total open string action invariant under the
gauge transformation of the $B$-field. Taking the exterior derivative of Eq.~\eqref{A.8} yields
\begin{equation}
F_{\mathrm{wv}}:=\mathrm dA_{\mathrm{wv}}
\longmapsto F_{\mathrm{wv}}-\mathrm d(\iota^\ast \mathbf\Lambda),
\label{A.10}
\end{equation}
whereas Eq.~\eqref{A.6} gives
\begin{equation}
\iota^\ast B\longmapsto\iota^\ast B+\mathrm d(\iota^\ast \mathbf\Lambda).
\label{A.11}
\end{equation}
The two variations cancel in the sum, so
\begin{equation}
\mathcal F:=\iota^\ast B+F_{\mathrm{wv}},\qquad \delta\mathcal F=0.
\label{A.12}
\end{equation}
This proves Eq.~\eqref{5.19.1}. It also explains the physical role of the endpoint term:
it both describes the coupling of the endpoint to the brane gauge connection and restores gauge invariance in the presence of a world-sheet boundary.
Gauge invariance requires the combination $\mathcal F$ in the boundary dynamics, but it does not require $F_{\mathrm{wv}}$ to be non-zero in every configuration.
In particular, $F_{\mathrm{wv}}=0$ is possible when the pullback of $B$ already supplies the required two-form.

\section{Variational derivation of the Neumann condition and the gluing block}
\label{app:B}

This appendix derives Eqs.~\eqref{5.20.1} and~\eqref{5.21.1} from the open string action \eqref{A.5}.
The bulk part of the first variation gives the equations of motion, whereas the boundary part determines the conditions obeyed by the string endpoints.
We use the boundary orientation and the convention $\partial_\pm=\partial_\tau\pm\partial_\sigma$ adopted in Section~\ref{Sec.V}.
Reversing the orientation of a boundary component changes the sign of its outward normal and correspondingly changes the local form of the boundary term;
the physical boundary prescription is unchanged when the orientation convention is treated consistently.

We first vary the metric part of the action. The terms containing derivatives of $\delta X^\mu$ are integrated by parts.
The resulting bulk terms contribute to the equations of motion, while the boundary term, with the chosen local orientation, is
\begin{equation}
\left.\delta S_G\right|_{\partial\Sigma}
=\frac{1}{2\pi\alpha'}\int_{\partial\Sigma}\mathrm d\tau\,
\delta X^\mu G_{\mu\nu}\partial_\sigma X^\nu.
\label{B.1}
\end{equation}
In coordinates adapted to the brane, the Dirichlet variations vanish, $\delta x^i=0$, and only the tangent variations are arbitrary.
Restricting the preceding expression to the Neumann directions therefore gives
\begin{equation}
\left.\delta S_G\right|_{\partial\Sigma,N}
=\frac{1}{2\pi\alpha'}\int_{\partial\Sigma}\mathrm d\tau\,
\delta x^m(G_N)_{mn}\partial_\sigma x^n.
\label{B.2}
\end{equation}

We next vary the bulk $B$-field coupling. For $H=\mathrm dB$, the variation of the pullback of a differential form can be written, by Cartan's formula, as
\begin{equation}
\delta(X^\ast B)=X^\ast (\iota_{\delta X}H) +\mathrm d\bigl[X^\ast (\iota_{\delta X}B)\bigr].
\label{B.3}
\end{equation}
The first term contributes to the bulk equations of motion. The second is an exact world-sheet form and, by Stokes' theorem, produces a boundary contribution. With the sign convention in Eq.~\eqref{A.5}, this contribution is
\begin{equation}
\left.\delta S_B\right|_{\partial\Sigma}
=-\frac{1}{2\pi\alpha'}\int_{\partial\Sigma}\mathrm d\tau\, \delta X^\mu B_{\mu\nu}\partial_\tau X^\nu. \label{B.4}
\end{equation}
Since the endpoint lies on the brane, the pullback to its tangent directions gives
\begin{equation}
\left.\delta S_B\right|_{\partial\Sigma,N}
=-\frac{1}{2\pi\alpha'}\int_{\partial\Sigma}\mathrm d\tau\, \delta x^m(\iota^*B)_{mn}\partial_\tau x^n. \label{B.5}
\end{equation}

It remains to vary the endpoint coupling to the world-volume gauge potential. From Eq.~\eqref{A.5}, this term is
\begin{equation}
S_A=-\frac{1}{2\pi\alpha'}\int_{\partial\Sigma}\mathrm d\tau\,
A_m(x)\partial_\tau x^m. \label{B.6}
\end{equation}
For clarity, omit the overall constant temporarily and define $I_A=\int\mathrm d\tau\,A_m(x)\dot x^m$.
The variation of $I_A$ is given by
\begin{equation}
\delta I_A=\int\mathrm d\tau\left[\partial_nA_m\,\delta x^n\dot x^m +A_m\partial_\tau(\delta x^m)\right].
\label{B.7}
\end{equation}
Integrating the second term by parts yields
\begin{equation}
\int\mathrm d\tau\,A_m\partial_\tau(\delta x^m)
=\left[A_m\delta x^m\right]_{\mathrm{corners}} -\int\mathrm d\tau\,(\partial_nA_m)\dot x^n\delta x^m.
\label{B.8}
\end{equation}
We take the variations to vanish at the temporal corners, so the corner term drops out. Relabelling dummy indices in the first term of Eq.~\eqref{B.7} and combining the two contributions then gives
\begin{align}
\delta I_A =\int\mathrm d\tau\,\delta x^m
\left(\partial_mA_n-\partial_nA_m\right)\dot x^n\notag =\int\mathrm d\tau\,\delta x^m(F_{\mathrm{wv}})_{mn}\partial_\tau x^n,
\label{B.9}
\end{align}
where $(F_{\mathrm{wv}})_{mn}=\partial_mA_n-\partial_nA_m$. Restoring the coefficient in Eq.~\eqref{B.6}, we find
\begin{equation}
\left.\delta S_A\right|_{\partial\Sigma}
=-\frac{1}{2\pi\alpha'}\int_{\partial\Sigma}\mathrm d\tau\,
\delta x^m(F_{\mathrm{wv}})_{mn}\partial_\tau x^n.
\label{B.10}
\end{equation}
The complete Neumann boundary variation is obtained by adding Eqs.~\eqref{B.2}, \eqref{B.5}, and~\eqref{B.10}.
Using $\mathcal F=\iota^*B+F_{\mathrm{wv}}$, this sum becomes
\begin{align}
\left.\delta S_{\mathrm{open}}\right|_{\partial\Sigma,N}
=\frac{1}{2\pi\alpha'}\int_{\partial\Sigma}\mathrm d\tau\,\delta x^m
\bigg[&(G_N)_{mn}\partial_\sigma x^n\notag -\mathcal F_{mn}\partial_\tau x^n\bigg]. \label{B.11}
\end{align}
The Neumann variations $\delta x^m$ are arbitrary at the boundary. Stationarity of the action therefore requires the coefficient of every such variation to vanish:
\begin{equation}
G_N\partial_\sigma x^m-\mathcal F\partial_\tau x^m=0.
\label{B.12}
\end{equation}
To express this condition in terms of the left- and right-moving derivatives, we use
\begin{equation}
\partial_\tau =\frac12(\partial_++\partial_-), \qquad \partial_\sigma =\frac12(\partial_+-\partial_-).
\label{B.13}
\end{equation}
Substitution into Eq.~\eqref{B.12} gives the light-cone Neumann condition quoted in the main text,
\begin{equation}
(G_N+\mathcal F)\partial_-x^m =(G_N-\mathcal F)\partial_+x^m.\label{B.15}
\end{equation}
Finally, the Neumann gluing matrix is defined by $\partial_-x^m=\mathbb R_N\partial_+x^m$. Inserting this definition into Eq.~\eqref{B.15} leads to
\begin{equation}
(G_N+\mathcal F)\mathbb R_N\partial_+x^m =(G_N-\mathcal F)\partial_+x^m. \label{B.16}
\end{equation}
Since this relation must hold for arbitrary allowed $\partial_+x^m$,
and assuming $G_N+\mathcal F$ is invertible on the Neumann subspace, we obtain
\begin{equation}
\mathbb R_N=(G_N+\mathcal F)^{-1}(G_N-\mathcal F),\label{B.17}
\end{equation}
which is Eq.~\eqref{5.21.1}. The inverse in this formula is only an inverse on the Neumann subspace; no inverse along Dirichlet directions is implied.
The same relation can be inverted to reconstruct the gauge-invariant two-form.
Multiplying Eq.~\eqref{B.17} from the left by $G_N+\mathcal F$ and then moving the terms containing $\mathcal F$ to the left and factoring it from the left yields
\begin{equation}
\mathcal F(\mathbb R_N+\mathbb I_N) =G_N(\mathbb I_N-\mathbb R_N).
\label{B.19}
\end{equation}
If $\mathbb I_N+\mathbb R_N$ is invertible, right multiplication by its inverse gives
\begin{equation}
\mathcal F=G_N(\mathbb I_N-\mathbb R_N) (\mathbb I_N+\mathbb R_N)^{-1}. \label{B.20}
\end{equation}
The order of matrix factors in Eq.~\eqref{B.20} follows from Eq.~\eqref{B.19} and must not be changed in general. Replacing the original fields and matrices by their dual counterparts gives the reconstruction formula used in Eq.~\eqref{5.24.1}.
\section{Structures on $D$-branes and boundary conditions in the Lie algebra frame}
\label{app.C}

In this appendix we investigate structures on $D$-branes and collect the relevant formulae for the boundary conditions in the Lie algebra frame.
We begin by defining a Dirichlet projector ${\mathbb{Q}}^{\mu}_{\;\nu}$ on the worldsheet boundary,
which projects the Dirichlet vectors onto the space normal to the brane. These vectors are eigenvectors of ${{\mathbb{R}}^{\mu}}_{\nu}$ with
eigenvalue $-1$. Hence, one can write the Dirichlet condition \eqref{5.1} in the following form
\begin{equation}\label{Dbrane3}
{\mathbb{Q}}^{\mu}_{\;\nu}~\partial_{\tau} X^{\nu} = 0.
\end{equation}
In addition, one may contract Eqs. \eqref{5.2} and \eqref{Dbrane3} to obtain
\begin{equation}\label{brane9}
{\mathbb{Q}}^{^{\mu}}_{_{\;\rho}}\;{{\mathbb{R}}^{^{\rho}}}_{{\;\nu}}\;=\;{{\mathbb{R}}^{^{\mu}}}_{_{\rho}}
\;{\mathbb{Q}}^{^{\rho}}_{_{\;\nu}}=-{\mathbb{Q}}^{\mu}_{\;\nu}.
\end{equation}
We also define a Neumann projector ${\mathbb{N}}^{\mu}_{\;\nu}$ which is complementary to ${\mathbb{Q}}^{\mu}_{\;\nu}$, namely,
\begin{equation}\label{brane10}
{\mathbb{N}}^{\mu}_{\;\nu}={\delta}^{\mu}_{\;\nu}-{\mathbb{Q}}^{\mu}_{\;\nu},~~~~~~~~~~~~~~{\mathbb{N}}^{\mu}_{\;\rho} ~ {\mathbb{Q}}^{\rho}_{\;\nu}=0.
\end{equation}
 The Neumann projector ${\mathbb{N}}^{\mu}_{\;\nu}$ projects vectors onto the tangent space of the $D$-brane,
whereas the Dirichlet projector ${\mathbb{Q}}^{\mu}_{\;\nu}$ projects onto the normal directions.
In adapted coordinates, the Dirichlet directions satisfy $ {\mathbb{R}}_D= - \mathbb{I}$.
However, the tangent directions are not necessarily eigenvectors of the full gluing matrix ${{\mathbb{R}}^{\mu}}_{\nu}$ with eigenvalue $+1$.
This property holds in the Neumann sector when the gauge-invariant two-form vanishes.
More generally, when a non-zero gauge-invariant two-form is present, the Neumann subspace is characterized by the boundary condition
involving the induced metric and the two-form, and its gluing matrix need not be the identity.
Accordingly, the Neumann projector identifies the tangent distribution, while the Neumann block of the gluing matrix
describes the boundary conditions within that distribution.
In the absence of an independent world-volume gauge field, the Neumann block is determined by the
pullback of the target space background matrix ${\cal E}_{_{{\mu \nu}}} = {G}_{_{{\mu \nu}}} + {B}_{_{{\mu \nu}}}$,
as expressed in Eq. \eqref{brane12}.
As shown in \cite{lind1,lind2}, the Neumann projector satisfies the following conditions
\begin{eqnarray}
\mathbb{N}}^\rho_{_{~\mu}}\; {\cal E}_{_{{\sigma \rho}}}\;{\mathbb{N}}^{\sigma}_{\;~\nu}- {\mathbb{N}}^\rho_{_{~\mu}}\; {\cal E}_{_{
\rho\sigma}}\;{\mathbb{N}}^{^{\sigma}}_{_{\;\lambda}}\;{{{\mathbb{R}}^{^{\lambda}}}_{{\nu}}&=&0, \label{brane12}\\
{{\mathbb{N}^\mu}_{_{{\rho}}}}\;
{G}_{{\mu \nu}}\;{{\mathbb{Q}}^{^{\nu}}}_{{\;\sigma}}&=&0, \label{brane12.1} \\
{{\mathbb{N}^\mu}_{_{{\gamma}}}}\;
{{\mathbb{N}^\rho}_{_{{\nu}}}}\; {{\mathbb{N}^\delta}_{_{{[\mu , \rho]}}}}  &=& 0.\label{brane12.111}
\end{eqnarray}
where ${{\mathbb{N}^\delta}_{_{{[\mu , \rho]}}}}$ in \eqref{brane12.111}, which serves as the integrability condition for the projector
${{{\mathbb{N}}^\mu}_{_{{\nu}}}}$ \cite{lind2}, is defined by:
\begin{eqnarray*}
{{\mathbb{N}^\delta}_{_{{[\mu , \rho]}}}}=\frac{1}{2} \big(\partial_\mu {{\mathbb{N}^\delta}_{{{\rho}}}} -\partial_\rho {{\mathbb{N}^\delta}_{{{\mu}}}}\big).
\end{eqnarray*}
The relation \eqref{brane12} is a condition on the NN part of ${{\mathbb{R}}^{^{\mu}}}_{{\;\nu}}$ stating
the definition of the $B$-field, and \eqref{brane12.1} implies the
diagonalization of the metric with respect to the $D$-brane.

Below, we shall write the above boundary conditions in the Lie algebra frame.
To this end, we use the right-invariant one-forms as mentioned in Section \ref{Sec.V}. Then, we have
\begin{eqnarray}
{\mathbb{Q}}^a_{~b}~ {\mathbb{R}}^b_{~c} = {\mathbb{R}}^a_{~b}~ {\mathbb{Q}}^b_{~c} &=& -{\mathbb{Q}}^a_{~c},\label{brane13.2}\\
{\mathbb{N}}^d_{{~a}}\; {E_{cd}}\;{\mathbb{N}}^{c}_{\;~b}-
{\mathbb{N}}^d_{_{~a}}\; { E_{ dc}}\;{\mathbb{N}}^{^{c}}_{_{\;e}}\;{{\mathbb{R}}^{e}}_{~b}&=&0,\label{brane13.3}\\
{{\mathbb{N}}^c_{~a}}\;
{\Omega}_{cd}\;{{\mathbb{Q}}^d}_{\;b}&=&0,\label{brane13.4}\\
{{\mathbb{N}}^c_{~a}}\;
{{\mathbb{N}}^e_{~b}}\; {{\mathbb{N}}}^d_{~[c , e]} &=& 0.\label{brane13.5}
\end{eqnarray}
In Eq. \eqref{brane13.4}, ${\Omega}_{ab}$ defined by ${\Omega}_{ab} = (R^{-1})^{\mu}_{~a}~ G_{\mu\nu}~(R^{-1})^{\nu}_{~b}$
is the Lie algebra metric.
As explained in Section \ref{Sec.V}, the condition \eqref{brane13.3} helps us to obtain the non-zero NN block of ${\mathbb{R}}$.
In addition, for a space-filling brane (along the Neumann directions),
from \eqref{brane13.3} one can get the gluing matrix in the form of \eqref{5.6}.

The same analysis can also be carried out for the dual geometry by replacing the original background fields,
gluing matrix, and projectors with their dual counterparts.
In particular, the dual Dirichlet and Neumann projectors are constructed from the dual gluing matrix,
and the corresponding dual boundary conditions are obtained by applying the dual versions of the relations presented above.
The integrability of the dual Neumann distribution must then be checked using the dual counterpart of condition \eqref{brane13.5},
while the Neumann boundary condition must be consistent with the induced dual metric and the gauge-invariant two-form on the dual D-brane.
This procedure is employed in Section \ref{Sec.V} to analyze the seven D-brane configurations related by the non-Abelian T-duality considered in this paper.


\begin{thebibliography}{99}

\bibitem{parinya1}
P. Karndumri,
Supersymmetric $AdS_6$ black holes from $ISO(3) \times U(1) F(4)$ gauged supergravity,
Eur. Phys. J. C {\bf 85} (2025) 158,
arXiv:2410.07837 [hep-th].

\bibitem{parinya2}
P. Karndumri,
Supersymmetric $AdS_2 \times \Sigma_2$  solutions from tri-sasakian truncation,
Eur. Phys. J. C {\bf 77} (2017) 689,  arXiv:1707.09633 [hep-th];
New supersymmetric $AdS_6$ black holes from matter-coupled $F(4)$ gauged supergravity,
Eur. Phys. J. Plus {\bf 139} (2024) 858, arXiv:2403.01746 [hep-th].

\bibitem{Minwoo.Suh1}
M. Suh,
Supersymmetric $AdS_6$ black holes from matter coupled $F(4)$ gauged supergravity,
J. High Energy Phys. {\bf 02} (2019) 108,
arXiv:1810.00675 [hep-th].

\bibitem{S.M.Hosseini}
S. M. Hosseini, K. Hristov, A. Passias and A. Zaffaroni,
6D attractors and black hole microstates,
J. High Energy Phys. {\bf 12} (2018) 001, arXiv:1809.10685 [hep-th].

\bibitem{Minwoo.Suh2}
M. Suh,
Supersymmetric $AdS_6$ black holes from  $F(4)$ gauged supergravity,
J. High Energy Phys. {\bf 01} (2019) 035, arXiv:1809.03517 [hep-th].

\bibitem{Klim1}
C. Klimcik  and P. Severa,
Dual non-Abelian duality and the Drinfeld double,
Phys. Lett. B {\bf 351} (1995) 455,
arXiv:hep-th/9502122.

\bibitem{Klim2}
C. Klimcik,
Poisson-Lie T-duality,
Nucl. Phys. (Proc. Suppl.)  B {\bf 46} (1996) 116,
arXiv:hep-th/9509095.

\bibitem{Buscher}
T. Buscher,
A symmetry of the string background field equations,
Phys. Lett. B {\bf 194} (1987) 59.


\bibitem{delaossa}
X. C. de la Ossa and F. Quevedo,
Duality symmetries from non-abelian isometries in string theory,
Nucl. Phys. B {\bf 403} (1993) 377, arXiv:hep-th/9210021;
A. Giveon and M. Rocek,
On nonabelian duality,
Nucl. Phys. B {\bf 421} (1994) 173, arXiv:hep-th/9308154;
E. Alvarez, L. Alvarez-Gaume and Y. Lozano,
On non-abelian duality,
Nucl. Phys. B {\bf 424} (1994) 155, arXiv:hep-th/9403155.


\bibitem{Drinfeld}
V. G. Drinfeld, {\it Quantum groups}, in  Proc. Intern. Cong. Math., Berkeley (1986) vol.
{\bf 1}, {Amer. Math. Soc.} (1987), pp. 798.


\bibitem{CC}
C. Albertsson and R. A. Reid-Edwards,
Worldsheet boundary conditions in  Poisson-Lie T-duality,
J. High Energy Phys. {\bf 03} (2007) 004,
arXiv:hep-th/0606024.

\bibitem{sfetsos}
K. Sfetsos,
Poisson-Lie T-duality and supersymmetry,
Nucl. Phys.  B (Proc. Suppl.) {\bf 56} (1997) 302,
arXiv:hep-th/9611199.

\bibitem{sfetsos1}
K. Sfetsos,
Canonical equivalence of non-isometric $\sigma$-models and Poisson-Lie T-duality,
Nucl. Phys.  B {\bf 517} (1998) 549,  	
arXiv:hep-th/9710163.

\bibitem{ooguri}
H. Ooguri, Y. Oz and Z. Yin,
D-branes on Calabi-Yau spaces and their mirrors,
Nucl. Phys. B {\bf 477} (1996) 407, hep-th/9606112;
K. Becker, M. Becker, D. R. Morrison, H. Ooguri, Y. Oz and Z. Yin,
Supersymmetric cycles in exceptional holonomy manifolds and Calabi-Yau 4-folds,
Nucl. Phys. B {\bf 480} (1996) 225, hep-th/9608116;
M. Kato and T. Okada, D-branes on group manifolds, Nucl. Phys. B {\bf 499} (1997) 583, hep-th/9612148;
S. Stanciu, D-branes in Kazama-Suzuki models, Nucl. Phys. B {\bf 526} (1998) 295, hep-th/9708166;
A. Recknagel and V. Schomerus, D-branes in Gepner models, Nucl. Phys. B {\bf 531} (1998) 185, hep-th/9712186.

\bibitem{Boundary.states}
C. Callan, C. Lovelace, C. Nappi and S. Yost,
Loop corrections to superstring equations of motion,
Nucl. Phys. B {\bf 308} (1988) 221;
M. Li,
Boundary states of D-branes and dy-strings,
Nucl. Phys. B {\bf 460} (1996) 351, hep-th/9510161;
C. G. Callan Jr. and I. R. Klebanov,
D-brane boundary state dynamics, Nucl. Phys. B {\bf 465} (1996) 473, hep-th/9511173.

\bibitem{Stanciu1}
S. Stanciu and A. Tseytlin,
D-branes in curved spacetime: Nappi-Witten background,
J. High Energy Phys. {\bf 06} (1998) 010, arXiv:hep-th/9805006.

\bibitem{Stanciu2}
J. M. Figueroa-O'Farrill and S. Stanciu,
More D-branes in the Nappi-Witten background,
J. High Energy Phys. {\bf 01} (2000) 024, arXiv:hep-th/9909164.

\bibitem{D.brane.WZW}
C. Klimcik and P. Severa,
Open strings and D-branes in WZNW model, Nucl. Phys. B {\bf 488} (1997) 653, arXiv:hep-th/9609112;
A. Yu. Alekseev and V. Schomerus,
D-branes in the WZW model, Phys. Rev. D {\bf 60} (1999) 061901, arXiv:hep-th/9812193;
S. Stanciu, D-branes in group manifolds, J. High Energy Phys. {\bf 01} (2000) 025, arXiv:hep-th/9909163;
U. Lindstrom and M. Zabzine,
D-branes in $N=2$ WZW models, Phys. Lett. B {\bf 560} (2003) 108, arXiv:hep-th/0212042;
R. Hernandez, G. Horcajada and F. Ruiz Ruiz,
D-branes with Lorentzian signature in the Nappi-Witten model,
J. High Energy Phys. {\bf 08} (2011) 047, arXiv:1104.4730 [hep-th].


\bibitem{eghbali11}
A. Eghbali and A. Rezaei-Aghdam,
Super Poisson-Lie symmetry of the $GL(1|1)$ WZNW model and worldsheet boundary conditions,
Nucl. Phys.  B {\bf 866} (2013) 26, arXiv:1207.2304 [hep-th];
Poisson Lie symmetry and D-branes in WZW model on the Heisenberg Lie group $H_4$,
Nucl. Phys. B {\bf 899} (2015) 165,
arXiv:1506.06233 [hep-th].

\bibitem{lind1}
C. Albertsson, U. Lindstrom and M. Zabzine,
$N=1$ supersymmetric sigma model with boundaries, I,
Commun. Math. Phys. {\bf 233} (2003) 403, arXiv:hep-th/0111161;
$N=1$ supersymmetric sigma model with boundaries, II,
Nucl. Phys. B {\bf 678} (2004) 295, arXiv:hep-th/0202069.

\bibitem{lind2}
C. Albertsson, U. Lindstrom and M. Zabzine,
Superconformal boundary conditions for the WZW model,  	
J. High Energy Phys. {\bf 05} (2003) 050, arXiv:hep-th/0304013.


\bibitem{lind3}
C. Albertsson, U. Lindstrom and M. Zabzine,
T-duality for the sigma model with boundaries,
J. High Energy Phys. {\bf 12} (2004) 056, arXiv:hep-th/0410217.

\bibitem{Snobl}
L. Hlavaty and L. Snobl,
Description of D-branes invariant under the Poisson-Lie T-plurality,
J. High Energy Phys. {\bf 07} (2008) 122, arXiv:0806.0963 [hep-th].

\bibitem{Tsimpis}
R. Terrisse, D. Tsimpis and C. A. Whiting,
D-branes and non-Abelian T-duality,
Nucl. Phys. B {\bf 947} (2019) 114733, arXiv:1811.05800 [hep-th].

\bibitem{Nakahara}
M. Nakahara, {\it Geometry, Topology and Physics},
2nd Edition, IOP, Bristol and Philadelphia (2003).

\bibitem{callan}
C. G. Callan, D. Friedan, E. Martinec and M. J. Perry,
String in background fields,
Nucl. Phys. B {\bf  262} (1985) 593.

\bibitem{c.hull}
C. M. Hull and K. Townsend,
String effective actions from sigma-model conformal anomalies,
Nucl. Phys. B {\bf 301} (1988) 197.

\bibitem{Metsaev}
R. Metsaev and A. Tseytlin,
Order $\alpha'$ (two-loop) equivalence of the string equations of motion and
the $\sigma$-model Weyl invariance conditions: Dependence on the dilaton and the antisymmetric tensor,
Nucl. Phys. B {\bf 293} (1987) 385.

\bibitem{Barrow}
J. D. Barrow and M. P. D\c{a}browski,
G\"{o}del universes in string theory,
Phys.  Rev. D {\bf 58} (1998) 103502.


\bibitem{Godel}
A. Eghbali, R. Naderi and A. Rezaei-Aghdam,
T-dualization of G\"{o}del string cosmologies via Poisson-Lie T-duality approach,
Eur. Phys. J. C {\bf 81} (2021) 68, arXiv:2002.00675.


\bibitem{Eghbali.2}
A. Eghbali,
Exact conformal field theories from  mutually T-dualizable $\sigma$-models,
Phys.  Rev. D {\bf 99} (2019) 026001,
arXiv:1812.07664 [hep-th].

\bibitem{ERA2}
A. Eghbali, R. Naderi and A. Rezaei-Aghdam,
Non-Abelian T-duality of $AdSd_{d \leq 3}$ families by Poisson-Lie T-duality,
Eur. Phys. J. C {\bf 82} (2022) 580,
arXiv:2111.07700 [hep-th].

\bibitem{egh.fortsch}
A. Eghbali, $AdS_3 \times S^3$ background from Poisson-Lie T-duality,
Fortsch. Phys. {\bf 72} (2024) 2400175, arXiv:2410.15053 [hep-th].

\bibitem{vonUnge}
R. von Unge,
Poisson-Lie T-plurality,
J. High Energy Phys. {\bf 07} (2002) 014,
arXiv:hep-th/0205245.

\bibitem{Arutyunov3}
G. Arutyunov, S. Frolov, B. Hoare, R. Roiban and A. A. Tseytlin,
Scale invariance of the $\eta$-deformed $AdS_5 \times S^5$ superstring, T-duality and modified type II equations,
Nucl. Phys. B {\bf 903} (2016) 262,
arXiv:1511.05795 [hep-th].

\bibitem{Wulff1}
R. Borsato and L. Wulff,
Target space supergeometry of $\eta $ and $\lambda$-deformed strings,
J. High Energy Phys. {\bf 10 } (2016) 045,
arXiv:1608.03570 [hep-th].

\bibitem{Wulff3.revised}
R. Borsato and L. Wulff,
Non-abelian T-duality and Yang-Baxter deformations of Green-Schwarz strings,
J. High Energy Phys. {\bf 08} (2018) 027,
arXiv:1806.04083 [hep-th].

\bibitem{samadi}
A. Eghbali, Y. Samadi and A. Rezaei-Aghdam,
Yang-Baxter deformations of the OSP$(1|2)$ WZW model,
Phys. Lett. B {\bf 872} (2026) 140062,
arXiv:2512.08315 [hep-th].

\bibitem{Bielli1}
D. Bielli, S. Penati, D. Sorokin and M. Wolf,
Super non-Abelian T-duality,
Nucl. Phys. B { \bf 983} (2022) 115904,
arXiv:112.12168 [hep-th].

\bibitem{Zwiebach}
B. Zwiebach, \emph{A First Course in String Theory}, Cambridge University Press, (2004).


\bibitem{Johnson}
C. V. Johnson,
D-brane primer, in Theoretical Advanced Study Institute in Elementary
Particle Physics (TASI 99):
Strings, Branes, and Gravity, {\bf 7} (2001) 129-350,
arXiv:hep-th/0007170.

\bibitem{Stanciu33}
S. Stanciu,
A note on D-branes in group manifolds: flux quantisation and D0-charge,
J. High Energy Phys. {\bf 10} (2000) 015, arXiv:hep-th/0006145.

\bibitem{Cheung}
Y. K. E. Cheung, L. Freidel,
Inner brane: A D3-brane in the Nappi-Witten model from an inner group automorphism,
Phys. Rev. D {\bf 79} (2009) 126007,
arXiv:0905.0540 [hep-th].


\end{thebibliography}
\end{document}